\documentclass[aps,prc,fleqn,superscriptaddress,floatfix,twocolumn]{revtex4}

\usepackage{graphicx,color}
\usepackage{amsmath,amssymb,amsfonts}

\newcommand{\be}{\begin{equation}}
\newcommand{\ee}{\end{equation}}
\newcommand{\ba}{\begin{eqnarray}}
\newcommand{\ea}{\end{eqnarray}}
\newcommand{\ban}{\begin{eqnarray*}}
\newcommand{\ean}{\end{eqnarray*}}
\newcommand{\nn}{\nonumber}

\begin{document}

\title{Electric Charge Separation in Strong Transient Magnetic Fields}
\medskip

\author{Masayuki Asakawa}
\affiliation{Department of Physics, Osaka University, Toyonaka 560-0043, Japan}

\author{Abhijit Majumder}
\affiliation{Department of Physics, The Ohio State University, Columbus, OH 43210,  USA}

\author{Berndt M\"uller}
\affiliation{Department of Physics, Duke University, Durham, NC 27708, USA}
\affiliation{Center for Theoretical \& Mathematical Sciences, Duke University, Durham, NC 27708, USA}


\begin{abstract}
We discuss various mechanisms for the creation of an asymmetric charge fluctuation with respect to the reaction plane among hadrons emitted in relativistic heavy-ion collisions. We show that such mechanisms exist in both, the hadronic gas and the partonic phases of QCD. The mechanisms considered here all require the presence of a strong magnetic field (the ``chiral magnetic effect''), but they do not involve parity or charge-parity violations. We analyze how a transient local electric current fluctuation generated by the chiral magnetic effect can dynamically evolve into an asymmetric charge distribution among the final-state hadrons in momentum space. We estimate the magnitude of the event-by-event fluctuations of the final-state charge asymmetry due to the partonic and hadronic mechanisms.
\end{abstract}

\maketitle

\section{Introduction}

Collisions of two heavy nuclei at high energy serve as a means for creating and exploring strongly interacting matter at the highest possible energy densities where matter is expected to assume the state of a quark-gluon plasma \cite{Harris:1996zx}. The properties of matter governed by the laws of quantum chromodynamics (QCD) have been studied in this way for a decade at the Relativistic Heavy Ion Collider (RHIC) at Brookhaven National Laboratory \cite{Arsene:2004fa,Adcox:2004mh,Back:2004je,Adams:2005dq}. The measurements performed in Au+Au, Cu+Cu, d+Au and p+p collisions at center-of-mass energies up to 200 GeV per nucleon pair have revealed several unusual properties of such superdense, strongly interacting matter \cite{Muller:2006ee}, most notably its very low kinematic shear viscosity and its high opacity with respect to energetic particles carrying a free color charge.

The strong interactions are known to respect space and time reflection symmetry to a very high degree. This is not a direct consequence of the laws of quantum chromodynamics which, in principle, permit a so-called $\theta$-term 
\be
{\cal L}_\theta = \frac{\theta}{32\pi^2} F^a_{\mu\nu} \tilde{F}^{a\mu\nu},
\ee
which violates time reversal symmetry. (Here $F^a_{\mu\nu}$ stands for the gluon field strength tensor and $\tilde{F}^{a\mu\nu}$ for its dual.) Rather, the symmetry conserving nature of QCD has been established by precise experiments that set limits on the intrinsic electric dipole moment of the neutron. The present experimental limit \cite{Baker:2006ts} implies that the coefficient of the possible CP violating term in the QCD Lagrangian $\theta < 0.7 \times10^{-11}$ \cite{Kim:2008hd}. The reason for its suppression or even its complete absence is not known; an often considered mechanism is the postulated existence of a new, spontaneously broken symmetry, called Peccei-Quinn symmetry \cite{Peccei:1977hh}, which would give rise to a new light, neutral particle, the axion \cite{Wilczek:1977pj}. 

Even if CP-symmetry is not violated in the normal QCD vacuum, it is conceivable that it is violated in an excited state or ``false'' QCD vacuum. Theorists have speculated that such a state might be created when QCD matter is heated above the critical temperature for confinement and chiral symmetry breaking and accidentally cools down into a configuration that is characterized by a nonzero value of $\theta$ \cite{Morley:1983wr,Kharzeev:1998kz}, and experimental observables capable of detecting the formation of symmetry violating domains of QCD matter in heavy ion collisions have been proposed \cite{Kharzeev:1999cz, Voloshin:2000xf,Finch:2001hs}.

Colliding nuclei create not only the highest temperatures, but also the strongest magnetic fields attainable in the laboratory \cite{Rafelski:1975rf}. For the purpose of studying the properties of QCD matter, the coherent magnetic field generated by the moving nuclei can be considered as ``external'': the value of the magnetic field at a given point is determined by the global charge distribution of the colliding nuclei and is thus, in good approximation, independent of the local strong interaction dynamics. However, the presence of the magnetic field can affect the interactions among the quarks and anti-quarks, which simultaneously carry electric and color charge.

Because magnetic fields are odd under time reversal (or equivalently, under a combined charge conjugation and parity (CP) transformation), the time reversal symmetry of a quantum system is broken in the presence of an external magnetic field. A magnetic field ${\bf B}$ can also combine with an electric field ${\bf E}$ to form the Lorentz invariant ${\bf E}\cdot{\bf B}$ which changes sign under a parity  transformation. An especially interesting aspect of magnetic fields is thus that they can be used to probe the response of strongly interacting matter in the pseudoscalar, CP-odd sector of QCD.  In the simultaneous presence of parallel electric and magnetic fields, QCD matter is thus dynamically forced to explore states with unnatural CP symmetry, eliminating the need to invoke a ``spontaneous'' symmetry violating transition.

As first pointed out by Kharzeev, McLerran, and Warringa \cite{Kharzeev:2007jp}, the coherent magnetic field generated by two heavy nuclei colliding off-centrally at high energy can convert topological charge fluctuations in the quantum chromodynamic (QCD) vacuum into global fluctuations around the electric charge symmetry with respect to the reaction plane. They called this mechanism the {\em chiral magnetic effect} and  interpreted it as a ``local'' violation of P and CP symmetry. The chiral magnetic effect was further analyzed by Fukushima, Kharzeev, and Warringa~\cite{Fukushima:2008xe,Kharzeev:2009pi,Fukushima:2009ft}, who argued that it is specific to the high temperature phase of QCD matter, commonly called the quark-gluon plasma, in which quarks are liberated from confinement into hadrons, because only then could the local electric current fluctuations be converted into charge fluctuations on the nuclear scale. The same conclusion was reached by Fu, Liu, and Wu \cite{Fu:2010pv}, who calculated the electric charge separation in a magnetic field in the framework of the Polyakov loop--Nambu--Jona-Lasinio (PNJL) model. The temperature dependence of the chiral magnetic effect and its dependence on the magnetic field in the presence of instanton fluctuations was investigated by Nam \cite{Nam:2009jb}.  Lattice QCD simulations of the chiral magnetic effect were reported by Buividovich {\em et al.} \cite{Buividovich:2009wi} (in quenched SU(2) gauge theory) and Abramczyk {\em et al.} \cite{Abramczyk:2009gb} (in $(2+1)$-flavor QED+QCD). A review of the chiral magnetic effect and its possible manifestation in relativistic heavy-ion collisions can be found in \cite{Kharzeev:2009fn}.

Inspired by these theoretical arguments and by experimental considerations related to the detection of parity violating effects in QCD \cite{Voloshin:2004vk}, the STAR collaboration has analyzed the final states of heavy-ion collisions at RHIC (Au+Au and Cu+Cu) for the presence of non-vanishing fluctuations in P- and CP-odd observables. The most promising of these is the asymmetry with respect to the reaction plane of the average emission angle between like-sign and opposite-sign hadrons defined by the event average
\be 
\label{eq-CSTAR}
C^{(\pm,\pm)} = \left\langle \cos\big(\phi_\alpha^{(\pm)} + \phi_\beta^{(\pm)} -2 \Psi_{\rm RP}\big) \right\rangle
\ee
where $\phi_\alpha,\phi_\beta$ denote the azimuthal emission angles of any pair of hadrons with respect to the beam axis, and $\Psi_{\rm RP}$ denotes the azimuthal orientation of the reaction plane. A difference
\be
\Delta Q \sim C^{(++)}+C^{(--)} - 2C^{(+-)}
\ee 
indicates a charge asymmetry of the kind suggested by the chiral magnetic effect. The existence of such an asymmetry was reported in a recent publication \cite{Abelev:2009uh}.

The purpose of our manuscript is to address two issues not clearly discussed in previous publications. First, we wish to point out that the mechanism producing an asymmetric charge fluctuation of the kind induced by the chiral magnetic effect is not constrained to an environment in which quarks are deconfined. We show that an analogous mechanism, {\em magnetic $\pi-\rho$ conversion}, exists in a confined hadronic environment. This should be expected in view of the general principle of {\em parton-hadron duality}, which is pervasive in QCD \cite{Dokshitzer:1991eq} and asserts that every mechanism present at the partonic level has an analogue in the world of hadrons. 

A second purpose of our manuscript is to analyze how a local electric charge asymmetry can be converted into a global asymmetry of charged hadron emission. This is a nontrivial problem, because the electric current fluctuations with respect to the reaction plane constitute an effect in position space, whereas the observable (\ref{eq-CSTAR}) measures an asymmetry in momentum space. We show that the presence of transverse collective flow and, more generally, the properties of hadronic freeze-out are essential to converting an asymmetry in position space into a momentum space asymmetry.

We begin by discussing the partonic and hadronic mechanisms for the electric current fluctuations underlying the chiral magnetic effect (CME) and then investigate the question of how a fluctuation in the electric current with respect to reaction plane gets converted into a fluctuation of the emission of charge hadrons. Next, we estimate the size of certain hadronic and partonic (gluonic) contributions to a charge asymmetry fluctuation. Our work concludes with a summary and outlook.

\section{Pseudoscalar Interactions of the Electromagnetic Field}

\subsection{Role of the Axial Anomaly}

In the normal QCD vacuum with its spontaneously broken chiral symmetry, the leading interaction involving the invariant ${\bf E}\cdot{\bf B}$ is \cite{Schwinger:1951nm,Feldmann:1999uf}:
\be 
\label{eq-EB-had}
{\cal L} = \sum_{i=\pi^0,\eta,\eta'} \frac{\alpha}{\pi f_i} \phi_i \, {\bf E}\cdot{\bf B} ,
\ee
where $\phi_i$ denotes one of the neutral pseudoscalar meson fields, $f_i$ is the respective meson decay constant, and $\alpha$ stands for the electromagnetic fine-structure constant. The interaction (\ref{eq-EB-had}) mediates the two-photon decays of the neutral pseudoscalar mesons. 

In the deconfined, chirally symmetric phase of QCD, the leading interaction is of the form
\be
\label{eq-EB-gg}
{\cal L}' = \kappa \alpha\alpha_s ({\bf E}^a\cdot{\bf B}^a) \, ({\bf E}\cdot{\bf B}) ,
\ee
where ${\bf E}^a$ and ${\bf B}^a$ denote the chromoelectric and chromomagnetic fields, respectively, and $\alpha_s=g^2/4\pi$ is the QCD coupling constant. The interaction (\ref{eq-EB-gg}) is not fundamental to QCD; it arises as an effective interaction at the one-quark-loop level \cite{Elze:1998wm}. 

Both interactions are closely related to the electromagnetic axial anomaly, which relates the divergence of the isovector axial current  to the pseudoscalar invariant of the electromagnetic field
\be
\label{eq-QED-an}
\partial_\mu j_5^{(3)\mu} = i(m_u \bar{u}\gamma_5 u + m_d \bar{d} \gamma_5 d)
 -  \frac{5N_c\alpha}{9\pi} {\bf E}\cdot{\bf B} ,
\ee
where $m_i$ are the current quark masses and $N_c=3$ is the number of colors. The flavor-singlet (isoscalar) axial current in QCD has the anomaly
\ba
\label{eq-QCD-an}
\partial_\mu j_5^{(0)\mu} &=& \sum_i 2i m_i \bar{q}_i \gamma_5 q_i 
 -  \frac{2N_c\alpha}{\pi}\sum_f\frac{e_f^2}{e^2}\, {\bf E}\cdot{\bf B}
 \nn \\
&& \qquad - \frac{N_f \alpha_s}{\pi}\, {\bf E}^a\cdot{\bf B}^a ,
\ea
where $N_f$ the number of light flavors, $e_f$ denotes the electric charge of each quark flavor, and $\alpha_s=g^2/4\pi$ is the QCD coupling constant. (For a recent review of the physics of the axial anomaly see \cite{Ioffe:2006ww}.) The pseudoscalar invariant of the gluon field is related to the topological charge density (or Chern-Simons number density) of the gluon field
\be
\label{eq-rhoW}
\rho_{\rm CS} = \frac{g^2}{32\pi^2} F^{a\mu\nu}\,\tilde{F}^a_{\mu\nu} = \frac{\alpha_s}{2\pi} {\bf E}^a\cdot{\bf B}^a .
\ee

\subsection{Anomalous Current in QCD}

The ``chiral magnetic effect'' occurs when the dynamics governing the nuclear reaction induces an electromagnetic current fluctuation $\delta{\bf j}$ that is parallel to the magnetic field ${\bf B}$ generated by the colliding nuclei. The electromagnetic current operator is given by
\be
j^\mu(x) = \frac{\delta}{\delta A_\mu(x)} \int d^4x\, {\cal L}_{\rm eff}[A] ,
\ee
where ${\cal L}_{\rm eff}[A]$ is the effective Lagrangian for the electromagnetic field. In order to induce a current {\em parallel} to the external magnetic field ${\bf B}$, ${\cal L}_{\rm eff}$ must contain a contribution of the pseudoscalar form 
\be
\label{eq-Leff}
{\cal L}_{\rm eff} = \frac{1}{4} {\cal P}\, F_{\mu\nu} \tilde{F}^{\mu\nu}
= {\cal P}\, {\bf E}\cdot{\bf B}  , 
\ee
where $\tilde{F}^{\mu\nu} = (1/2) \varepsilon^{\mu\nu\alpha\beta} F_{\alpha\beta}$ is the dual field tensor and ${\cal P}$ is a pseudoscalar operator constructed from strongly interacting fields. Using this form of the effective Lagrangian, one obtains the following general form for the anomalous electromagnetic current:
\be
\label{j-an}
j^\mu_{\rm an}(x) = -  (\partial_\nu {\cal P})\, \tilde{F}^{\mu\nu} ,
\ee
which was first derived in general form by D'Hoker and Goldstone \cite{DHoker:1985yb}.  By virtue of Maxwell's equations and the antisymmetry of $\tilde{F}^{\mu\nu}$, one easily confirms that the anomalous current is conserved:
\be
\partial_\mu j^\mu_{\rm an}(x) = -  (\partial_\mu\partial_\nu{\cal P})\, \tilde{F}^{\mu\nu}
- (\partial_\nu {\cal P})\, \partial_\mu\tilde{F}^{\mu\nu} = 0 .
\ee

\begin{center}
\begin{figure}[!htb]
\includegraphics[height=3cm]{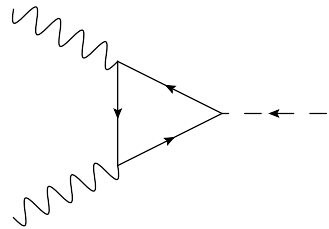}
\caption{Feynman diagram describing the anomalous coupling between two photons and a neutral pseudoscalar meson via a triangular quark loop.}
\label{fig-pigamgam}
\end{figure}
\end{center}

\begin{center}
\begin{figure}[!htb]
\includegraphics[height=4cm]{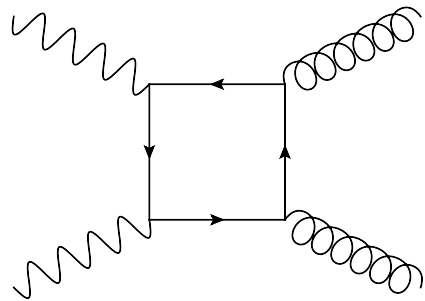}
\caption{Feynman diagram describing the effective pseudoscalar coupling between two gluons and two photons via a quark loop.}
\label{fig-box}
\end{figure}
\end{center}

As mentioned in the Introduction, the form of possible pseudoscalar operators ${\cal P}$ differs fundamentally between the phases of QCD with and without spontaneous chiral symmetry breaking. In the normal QCD vacuum with spontaneously broken chiral symmetry, relevant pseudoscalar operators exist in the form of neutral pseudoscalar mesons ($i = \pi^0, \eta, \eta'$), which give rise to interactions of the form (\ref{eq-Leff}) via the Adler--Bell--Jackiw triangle anomaly (see Fig.~\ref{fig-pigamgam}):
\be
\label{eq-Leff-had}
{\cal L}_{\rm eff}^{\rm (had)} = \sum_i \frac{\alpha}{\pi f_i}\phi_i\, ({\bf E}\cdot{\bf B}) ,
\ee
where $\phi_i$ denotes the pseudoscalar meson fields and $f_i$ stands for the meson decay constant. In the chirally unbroken, deconfined phase of QCD the relevant pseudoscalar operator is given by the topological charge (or Chern-Simons number) density $\rho_{\rm CS}$ defined in (\ref{eq-rhoW}). The combined effective QED-QCD action for soft gauge fields contains a term of the form \cite{Elze:1998wm}
\be
\label{eq-Leff-QCD}
{\cal L}_{\rm eff}^{\rm (QECD)} = \kappa\, \alpha\alpha_s\, ({\bf E^a}\cdot{\bf B^a})\,({\bf E}\cdot{\bf B}) .
\ee
The coupling coefficient $\kappa \propto \sum_f (e_f/e)^2$, where $e_f$ denotes the electric charge of a quark of flavor $f$, is given by the quark box diagram (see Fig.~\ref{fig-box}) with two gluon and two photon vertices. 

In the limit that the magnetic field, the current, and the exchanged gluons are ``soft'' and the quark mass $m$ is large, the quark box diagram is obtained from the generalization of the Heisenberg-Euler effective Lagrangian of QED to include both, electromagnetic and Yang-Mills fields. At fourth order in the fields, the effective Lagrangian in the vacuum reads \cite{Elze:1998wm}:
\be
\label{LQECD}
{\cal L}_{\rm QECD}^{(4)} = \frac{-1}{360\pi^2} {\rm Tr}_{C,F} 
\int_0^\infty s\,ds\, e^{im^2s} (4\hat{\cal F}^2 + 7\hat{\cal G}^2) ,
\ee
where 
\be
\hat{\cal F} = \frac{1}{4} \hat{F}_{\mu\nu} \hat{F}^{\mu\nu} , \qquad
\hat{\cal G} = \frac{1}{8} \varepsilon^{\mu\nu\alpha\beta} \hat{F}_{\mu\nu} \hat{F}_{\alpha\beta}
\ee
are the two Lorentz invariants of the ``mixed'' field tensor
\be
\hat{F}_{\mu\nu} = e F_{\mu\nu} + g F_{\mu\nu}^a t^a .
\ee
Here $F_{\mu\nu}$ denotes the electromagnetic field tensor, $F_{\mu\nu}^a$ denotes the field tensor of the color field, and the Gell-Mann matrices $t^a$ are generators of color-SU(3) in the fundamental representation. The trace in (\ref{LQECD}) runs over colors and quark flavors; the relevant color trace is ${\rm Tr}(t^at^b) = \delta^{ab}/2$. The contribution of interest to us is
\ba
{\rm Tr}_C (\hat{\cal G}^2) 
&=& \frac{1}{16}  e^2g^2 (\tilde{F}^a_{\mu\nu} F^{a\mu\nu}) (\tilde{F}_{\mu\nu} F^{\mu\nu}) + \cdots
\nn \\
&=& e^2g^2 ({\bf E}^a\cdot{\bf B}^a) ({\bf E}\cdot{\bf B}) + \cdots .
\ea
The effective vacuum action arising from the quark box diagram in Fig.~\ref{fig-box} is thus:
\be
{\cal L}_{{\rm eff},0}^{\rm (QECD)} 
= \frac{14}{45}\alpha\alpha_s \sum_f \frac{(e_f/e)^2}{m_f^4}\, ({\bf E}^a\cdot{\bf B}^a) ({\bf E}\cdot{\bf B}) .
\ee
It describes the contribution of a heavy quark flavor ($m_f \gg \Lambda_{\rm QCD}$) to the pseudoscalar effective gluon-photon coupling.

The finite temperature effective QED action has been studied repeatedly \cite{Elmfors:1994fw,Elmfors:1998ee,Gies:1998vt}. A detailed account of the weak-field expansion of the thermal effective action can be found in \cite{Elmfors:1998ee}. The expression for the mixed QED--QCD pseudoscalar contribution differs from the pure QED result in the same way as described above for the vacuum case. The explicit expression, after a Wick rotation in the integration variable $s$, is
\ba
{\cal L}_{{\rm eff},T}^{\rm (QECD)} 
& = & \frac{7}{90\,\pi^{3/2}}\,{\rm Tr}_{C,F} (\hat{\cal G}^2) \, \int_0^\infty s^{3/2}ds\,e^{-m^2s}
\nn \\
& & \qquad\qquad \times T \sum_{n=0}^\infty e^{-[(2n+1)\pi T]^2s} .
\ea
The integral is easily carried out, and we obtain as final result for the coefficient $\kappa$ defined in (\ref{eq-Leff-QCD}) in the high temperature limit ($T\gg m$):
\be
\label{kappa-T}
\kappa(T) \approx  \frac{651\,\zeta(5)}{720\,(\pi T)^4}\, \sum_f  (e_f/e)^2 
= \frac{217\,\zeta(5)}{360\,(\pi T)^4} ,
\ee
where the last form includes the contributions from the three light quark flavors. We note that the expression is independent of the quark masses as long as these are small compared to the temperature.

\begin{center}
\begin{figure}[!htb]
\includegraphics[width=0.9\linewidth]{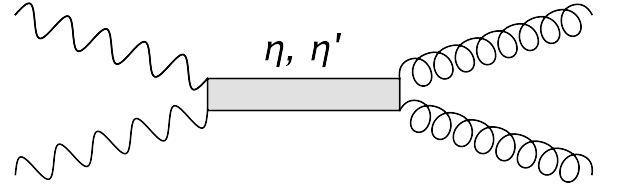}
\caption{Feynman diagram describing the effective pseudoscalar coupling between two gluons and two photons via a virtual $\eta$- or $\eta'$-meson.}
\label{fig-eta-gg}
\end{figure}
\end{center}

\begin{center}
\begin{figure}[!htb]
\includegraphics[width=0.9\linewidth]{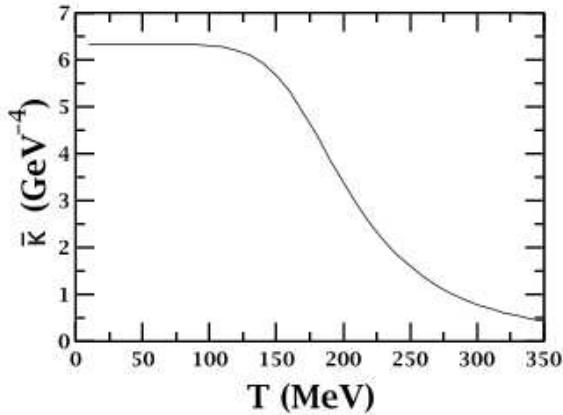}
\caption{Temperature dependence of the coefficient $\kappa$ in the effective pseudoscalar QED--QCD interaction (\ref{eq-Leff-QCD}).}
\label{fig-kappa}
\end{figure}
\end{center}

For the three light quark flavors, the interaction between gluons and photons in the pseudoscalar channel is intermediated by the iso-singlet pseudoscalar mesons $\eta$ and $\eta'$ (see Fig.~\ref{fig-eta-gg}). The effective interactions are determined by the electromagnetic and chromodynamic anomaly, respectively \cite{Ali:2000ci}:
\ba
\label{eta-anom-em}
{\cal L}_{\eta\gamma\gamma} &=& \frac{\alpha}{\pi f_\eta}
\left(\frac{\cos\theta}{\sqrt{6}} - \frac{2}{\sqrt{3}}\sin\theta \right) \eta \, ({\bf E}\cdot{\bf B}) ;
\\
\label{etap-anom-em}
{\cal L}_{\eta'\gamma\gamma} &=& \frac{\alpha}{\pi f_\eta}
\left(\frac{2}{\sqrt{3}}\cos\theta + \frac{\sin\theta}{\sqrt{6}} \right) \eta' \, ({\bf E}\cdot{\bf B}) ;
\\
\label{eta-anom-qcd}
{\cal L}_{\eta gg} &=& - \frac{\alpha_s}{2\pi f_\eta} \sqrt{3}\sin\theta\, \eta \, ({\bf E}^a\cdot{\bf B}^a) ;
\\
\label{etap-anom-qcd}
{\cal L}_{\eta' gg} &=& \frac{\alpha_s}{2\pi f_\eta} \sqrt{3}\cos\theta\, \eta' \, ({\bf E}^a\cdot{\bf B}^a) .
\ea
Here $\theta \approx -20^{\rm o}$ is the flavor singlet-octet mixing angle and $f_\eta \approx 1.2\, f_\pi \approx 157$ MeV \cite{Feldmann:1999uf}. The effective pseudoscalar photon-gluon interaction is then, in the low-energy limit, given by:
\ba
{\cal L}_{{\rm eff},0}^{\rm (QECD)} 
&=& \left[ 1+ \frac{\tan\theta}{2\sqrt{2}} - \frac{m_{\eta'}^2}{m_\eta^2}
\left( \frac{\tan\theta}{2\sqrt{2}} - \tan^2\theta \right) \right]
\nn \\
&& \times \frac{\alpha\alpha_s \cos^2\theta}{\pi^2 f_\eta^2 m_{\eta'}^2}\,
({\bf E}^a\cdot{\bf B}^a) ({\bf E}\cdot{\bf B}) 
\ea
corresponding to the zero-temperature coefficient
\be
\label{kappa-0}
\kappa_0 \approx \frac{1.46}{\pi^2 f_\eta^2 m_{\eta'}^2} .
\ee
A useful form that interpolates between the low- and high-temperature limits (\ref{kappa-0}, \ref{kappa-T}) is:
\be
\bar{\kappa}(T) \approx \frac{\kappa_0\, \kappa(T)}{\sqrt{\kappa_0^2 + \kappa(T)^2}}
\ee
The temperature dependence of $\bar{\kappa}(T)$ is shown in Fig.~\ref{fig-kappa}.

\subsection{Hadronic Current}

We first consider the hadronic current in more detail, starting from the general form of the ``anomalous'' interaction (\ref{eq-Leff}). Invoking the general form (\ref{j-an}) of the anomalous current, we obtain for the vector current density in vectorial notation:
\be 
\label{eq-J-pi}
{\bf j}_{\rm an}^{(\pi)} = \left[ (\partial_t{\cal P})\, {\bf B} + (\nabla{\cal P})\times{\bf E} \right] .
\ee
Identifying ${\cal P} = (\alpha/\pi f_\pi)\phi$, we obtain for the anomalous current induced by a magnetic field acting on the pion field:
\be
{\bf j}_{\rm an}^{(\pi)} = \frac{\alpha}{\pi f_\pi} \dot\phi \, {\bf B} ,
\ee
where the dot, as usual, indicates a time derivative.  What may seem strange about this result is that a neutral meson field (the neutral pion field) combines with the electromagnetic field to generate an electric current. However, the effect is easily understood when one recognizes that an external electromagnetic field deflects the quark and antiquark constituents of the pion in opposite directions causing the neutral pion to become internally polarized. The current (\ref{eq-J-pi}) thus describes the polarization current. Before proceeding we note that the magnetic field generated by two heavy nuclei colliding at relativistic energies is extremely large, and its strength, as measured by the product $eB$, can be of order $m_\pi^2$ (see Fig.~A1 in \cite{Kharzeev:2007jp}).

\begin{center}
\begin{figure}[!htb]
\includegraphics[height=3cm]{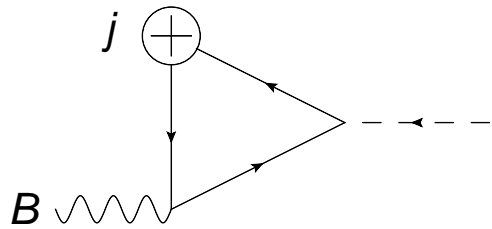}
\caption{Quark triangle diagram for the electromagnetic current ${\bf j}$ induced by the presence of a magnetic field $B$ in a neutral pion. The cross at the upper corner of the triangle indicates the insertion of the electromagnetic current operator $e_f\gamma^\mu$.}
\label{fig-tri}
\end{figure}
\end{center}

\begin{center}
\begin{figure}[!htb]
\includegraphics[height=2.5cm]{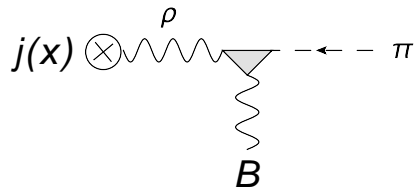}
\caption{Feynman diagram for the electromagnetic current ${\bf j}$ induced by the conversion of a neutral pion into a neutral $\rho$-meson in the presence of a magnetic field $B$. In the vector dominance model, the electromagnetic current $j^\mu$ is proportional to the $\rho$-meson field, see Eq.~(\ref{eq:VMD}).}
\label{fig-tri-rho}
\end{figure}
\end{center}

The polarization current (\ref{eq-J-pi}) is highly localized and confined to the interior of the neutral pion. It is this current that generates the electric field vector of one of the two photons in the decay of the neutral pion. However, in order to generate a macroscopic current that can be observed as a fluctuating charge asymmetry with respect to the reaction plane, we need an effective coupling to charged hadrons that can move freely through the nuclear fireball. The relevant Feynman diagram is shown in Fig.~\ref{fig-tri-rho}. Here the magnetic field excites the neutral pion into a neutral vector meson, e.g. the $\rho^0$-meson, which is polarized in the direction of the magnetic field and induces a polarization current among the charged pions in the medium. 

The effective $\pi^0V\gamma$ coupling, where $V$ stands for any neutral vector meson, has been studied extensively \cite{O'Donnell:1981sj}. Experimentally, the strengths of the couplings are known from the partial decay widths of the vector mesons: 
$\Gamma_{\omega \to \pi^0\gamma} = 0.79$ MeV, $\Gamma_{\rho^0 \to \pi^0\gamma} = 0.077$ MeV 
\cite{Achasov:2003ed}. These observations imply that there exists an effective interaction of the form \cite{Zhu:1998bm}
\be 
\label{eq-Vpigamma}
{\cal L}_{\rm eff}^{(V\pi\gamma)} = \frac{e\, g_{V\pi\gamma}}{2m_V} \varepsilon_{\mu\nu\alpha\beta} 
  F^{\mu\nu}  (\partial^\alpha \phi) V^\beta,
\ee
where $\phi$ denotes the neutral pion field and $V^\mu$ the neutral vector meson field ($V=\omega,\rho^0$). The effective coupling constant $g_{\rho\pi\gamma} = 0.58$ \cite{Zhu:1998bm}. For later purposes, we also note the strength of the $\rho\eta'\gamma$ coupling, $g_{\rho\eta'\gamma} \approx 1.31$ \cite{Feldmann:1999uf}, which can be deduced from the measured decay width
$\Gamma_{\eta'\to\rho\gamma} = \alpha \, g_{\rho\eta'\gamma}^2 \, p_{\rm cm}^3/m_{\eta'}^2 
\approx 60\pm 5 {\rm ~keV}$  \cite{Amsler:2008zzb}.

Later we will evaluate the fluctuation of the magnetically induced electric current $\langle j_i(x) j_k(x')\rangle$ in a thermal hadron gas. The calculation is much simplified by noting that the vector meson dominance (VMD) model allows us to replace the hadronic electromagnetic current operator $j^\mu$ by the vector meson operator:
\be
\label{eq:VMD}
j^\mu = - \frac{e\, m_\rho^2}{g_{\rho}} \,\rho^\mu .
\ee
For the sake of simplicity, we focus on the $\rho$-meson contribution, because only the $\rho$-meson couples to the two-pion channel, and pions are by far the most abundant hadrons in baryon-symmetric, thermal hadronic matter. Using the effective interaction (\ref{eq-Vpigamma})
\be 
{\cal L}_{\rm eff}^{(\rho\pi\gamma)} = \frac{e\, g_{\rho\pi\gamma}}{m_\rho} ({\bf B}\cdot\rho) \dot\phi ,
\ee
we obtain
\ba
\langle j_i(x) j_k(x') \rangle &=& e^2 m_\rho^2\, \frac{g_{\rho\pi\gamma}^2}{g_{\rho}^2}
 \sum_{mn} \int dy\, dy'\, 
\nn \\
& \times & \langle \rho_i(x) \rho_m(y) \rangle\, \langle \rho_n(y') \rho_k(x') \rangle
\nn \\
& \times & \langle \dot\phi(y) \dot\phi(y') \rangle \, B_m(y) B_n(y') .
\ea

Since we are interested in the matter contribution to the in-medium $\rho$-meson propagator $\langle \rho^\mu(x) \rho^\nu(y) \rangle$, which was calculated by Gale and Kapusta \cite{Gale:1990pn}, we write the position-space propagator as
\be 
\langle \rho^\mu(x) \rho^\nu(y) \rangle = \int \frac{d^4k}{(2\pi)^4} e^{ik(x-y)} D_\rho^{\mu\nu}(k) .
\ee
The momentum-space propagator (in the $\lambda=\infty$ gauge) can be separated into its longitudinal and transverse components:
\ba
\label{eq-Drho}
D_\rho^{\mu\nu}(k) &=& - \frac{P_{\rm L}^{\mu\nu}}{k^2-m_\rho^2-\Pi_{\rm L}(k)}
\nn \\
&&  - \frac{P_{\rm T}^{\mu\nu}}{k^2-m_\rho^2-\Pi_{\rm T}(k)} - \frac{k^\mu k^\nu}{m_\rho^2 k^2} .
\ea
$\Pi_{\rm L/T}(k)$ denotes the longitudinal and transverse components of the $\rho$-meson self-energy. 
Since we are primarily interested in slowly varying contributions to the induced current, i.e.\ in the limit $(k^0,{\bf k}) \ll m_\rho$, the matter contribution to (\ref{eq-Drho}) can be approximated as
\be
D_{\rho,\rm mat}^{\mu\nu}(k) = - \frac{\Pi_{\rm L}^{\rm (mat)}(k)}{m_\rho^4} P_{\rm L}^{\mu\nu}
  - \frac{\Pi_{\rm T}^{\rm (mat)}(k)}{m_\rho^4} P_{\rm T}^{\mu\nu} .
\ee
Explicit expressions for the thermal pion gas contributions to the $\rho$-meson self-energy can be found in \cite{Gale:1990pn}.

\subsection{Gluon Induced Current}

\begin{center}
\begin{figure}[!htb]
\includegraphics[height=2.5cm]{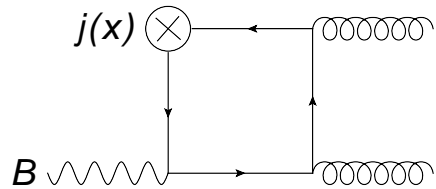}
\caption{Feynman diagram for the electromagnetic current ${\bf j}$ induced by two gluons in the pseudoscalar channel in the presence of an external magnetic field ${\bf B}$.}
\label{fig-J-gg}
\end{figure}
\end{center}

The diagram in Fig.~\ref{fig-J-gg}, on the other hand, represents a process whereby two gluons induce a current in the medium by coupling to a quark-antiquark pair in the presence of the strong magnetic field. Again invoking the general form of the anomalous electromagnetic current (\ref{j-an}), the anomalous current due to a magnetic field takes the following form for the pseudoscalar two-gluon coupling:
\be 
\label{eq-J-gg}
{\bf j}_{\rm an}^{(gg)} = \kappa\, \alpha\alpha_s\, \partial_t({\bf E}^a\cdot{\bf B}^a) {\bf B} .
\ee

\begin{center}
\begin{figure}[!htb]
\includegraphics[height=3.5cm]{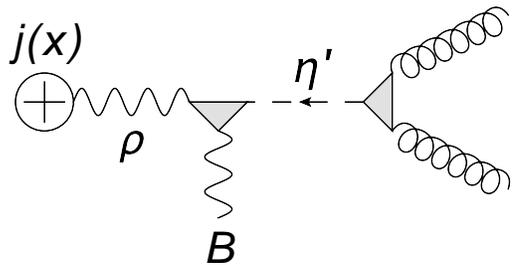}
\caption{Effective hadronic Feynman diagram for the anomalous electric current ${\bf j}$ induced by two gluons in the pseudoscalar channel in the presence of a magnetic field ${\bf B}$.}
\label{fig-ggetabj}
\end{figure}
\end{center}

As discussed in Sect. II.A, the pseudoscalar coupling of two gluons to the light quark sector in the vacuum is dominated by the $\eta'$ (or $\eta$) meson. An effective interaction analogous to (\ref{eq-Vpigamma}) then allows the $\eta'$ meson to convert into a $\rho^0$ meson in the presence of an external magnetic field. The induced anomalous electric current is then simply given by the VMD relation (\ref{eq:VMD}). The Feynman diagram describing this process is shown in Fig.~\ref{fig-ggetabj}.

\section{Electric Charge Asymmetry}

\subsection{Scenarios for the Reaction Plane Charge Asymmetry}

We now discuss five different scenarios that may contribute to the creation of an event-by-event reaction plane charge asymmetry in heavy ion collisions due to the interaction of the magnetic field with the pseudoscalar sector of the QCD matter. An overall guidance principle for assessing their relative importance is that the current fluctuations are proportional to the square of the magnetic field ${\bf B}$ acting on the strongly interacting particles. As analyzed by Kharzeev {\em et al.} \cite{Kharzeev:2007jp}, the coherent magnetic field generated by the moving charge density of the colliding nuclei in the center-of-mass reference frame is strongly peaked at the initial moment of impact and then falls off rapidly with an approximate power-law tail $|{\bf B}(\tau)| \propto \tau^{-2}$. As a consequence, scenarios occurring early in the reaction are strongly favored to be fertile environments for the creation of a charge asymmetry.

{\em Color Glass Condensate (CGC) Scenario:}
The magnetic field strength argument suggests that we should first and foremost consider the contribution to ${\bf E}^a\cdot{\bf B}^a$ from gluons contained in the saturated gluon wave functions ({\em i.e.} the color-glass condensate) of the colliding nuclei. These wave functions contain gluons of all colors and helicities and thus provide a bountiful supply of initial states containing a nonzero density of topological charge density $n_{\rm top}$ \cite{Lappi:2006fp}. Indeed, Shuryak and Zahed \cite{Shuryak:2002qz} have argued that the interaction of the colliding saturated gluon fields will create color field configuration of the sphaleron type, i.e.\ configurations carrying half-integer winding number, which then decay into multiple quark-antiquark pairs. 

{\em Quark-Gluon Plasma (QGP) Scenario:}
As already pointed out, the operator ${\bf E}^a\cdot{\bf B}^a$ is related to the Chern-Simons number density $\rho_{\rm CS}$ of the nonabelian gauge field. In thermal equilibrium, $\rho_{\rm CS}$ is given by the thermal winding number fluctuations in the quark-gluon plasma. This is the process considered by the authors of refs.~\cite{Kharzeev:2007jp,Fukushima:2008xe}.

{\em Glasma Scenario:}
Contributions to the anomalous current can also arise from topological charge density fluctuations occurring during the pre-equilibrium ``glasma'' phase. The magnitude of these fluctuations has been estimated by Kharzeev {\em et al.} \cite{Kharzeev:2001ev}. When glasma-phase quark pair production \cite{Gelis:2004jp,Gelis:2005pb} occurs in the presence of a strong, oriented magnetic field, the produced pairs will carry a nonzero electric current in the direction of the magnetic field. The generation of an anomalous current due to quark-pair production by a chromo-electric flux tube was recently investigated by Fukushima {\em et al.} \cite{Fukushima:2010vw} in the framework of the effective QCD+QED Lagrangian. This process also occurs at very early times, of the order of $Q_s^{-1}$, and thus shortly after the peak in the magnetic field strength.

{\em Corona Scenario:}
The surface region of the nuclear reaction zone never reaches the energy density required to form a quark-gluon plasma and thus remains in the hadronic phase of QCD matter throughout the reaction. Fluctuations of the charge asymmetry in this region will be produced by $\pi-\rho$ conversion in the magnetic field. This process can act on pions produced during the initial impact or even on virtual pions contained in the initial nuclear wave functions.

{\em Hadron Gas Scenario:}
After re-hadronization of the quark-gluon plasma until final freeze-out, the hot QCD matter proceeds through a thermal hadronic gas phase, where the magnetic $\pi-\rho$ conversion mechanism can operate throughout the bulk of the fireball. This scenario is disfavored by the lateness of its occurrence, as well as by its relatively short duration due to the rapid transverse expansion of the fireball at late times.

\subsection{Kinematic Considerations}

\begin{center}
\begin{figure}[!htb]
\includegraphics[width=0.9\linewidth]{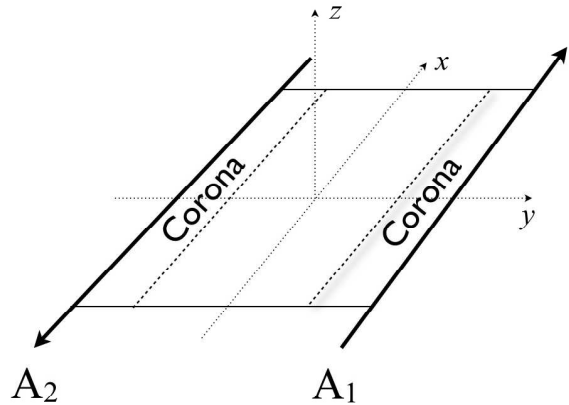}
\caption{Schematic view of the collision plane of the two heavy ions and the choice of our coordinate system.}
\label{react-plane}
\end{figure}
\end{center}

We first outline the strategy for the calculation of the charge asymmetry $\Delta Q$ with respect to the collision plane. We assume that the collision axis is the $x$-axis and the collision plane the $x-y$-plane, the axis perpendicular to the reaction plane is the $z$-direction. The strength of the magnetic field generated by the fast moving nuclei acting on the region of space-time causally connected to the mid-rapidity region is highly peaked around the collision moment $\tau=0$. This means that the current $j^\mu_{\rm an}(x)$ separating charges with respect to the collision plane is concentrated at early times, when the collective transverse flow of the matter is small. We can thus neglect (in first approximation) any collective flow during the period when the charge asymmetry is created. If there were no flow later on, the charge asymmetry would subsequently be erased by diffusion. Hence, the collective transverse flow which transports the separated charge further away f!
rom the collision plane is an essential component of the mechanism. The final charge asymmetry of all hadrons emitted above and below the collision plane will be determined by the charge distribution on the freeze-out hypersurface $\Sigma_f$. We will assume that the local net charge density $\rho(x)$ of the matter at a given point $x \in \Sigma_f$ can be parametrized by an effective local chemical potential $\mu_Q(x)$ for electric charge. Finally, it is important to keep in mind that the average charge asymmetry $\langle \Delta Q \rangle$ with respect to the collision plane is zero, and only the fluctuation $\langle (\Delta Q)^2 \rangle$ of this quantity can be measured by an event average.

The experiment measures fluctuations of the charge asymmetry with respect to the collision plane in momentum space, and the chiral magnetic effect creates a fluctuating charge density asymmetry with respect to the collision place in position space. In order to convert the latter into the former, it is necessary for a correlation to exist between the location of the emission point of a hadron and its angular distribution. Such a correlation can have two origins: collective transverse flow or the spatial orientation of the freeze-out hypersurface. In general, one expects both mechanisms to contribute; only for isochronous freeze-out or in the absence of collective transverse flow would a single mechanism exist. Below, we will estimate the amount of expected charge asymmetry with respect to the collision plane in these two extreme cases, which should give a lower bound on the real magnitude of the effect.

We also note that there are two space-time regions of the reaction volume that must be treated separately. One is the core region, where the initial energy density is high enough to create matter in the deconfined phase. Here the chiral magnetic effect induces a current spike, which leads to local charge separation as discussed above, followed by transport through the medium up to the final freeze-out surface. The other region is the ``corona'', where the energy density never exceeds the critical energy density required to create a quark-gluon plasma. Here the evolution proceeds entirely through the hadronic phase, and the axial vector current is always carried by hadronic interactions. Since there is anomaly matching \cite{tHooft} between the hadronic and quark-gluon phases of QCD, the anomalous contribution to the induced electromagnetic current remains unaffected, but the axial vector correlator will have a different representation, in terms of hadronic states.

Due to the Lorentz contraction, the magnetic field generated by the colliding nuclei in the central rapidity region is sharply peaked around the collision moment $\tau = 0$ and falls off by about two orders of magnitude during the first fm/c \cite{Kharzeev:2007jp}.  This is the time where most of the charge separation will be caused by the chiral magnetic effect. During this period, the transverse collective  flow is small, and we will therefore neglect collective flow during the charge separation phase. We then just need to calculate the net charge density created within the hot QCD medium during the period when the magnetic field is strong.  We will later discuss how the charge density created in the separation process is transported by the collective flow field and eventually freezes out into detected particles.

\subsection{Asymmetry Generated at Freeze-out}

\subsubsection{General formulation}

In order to calculate fluctuations in the charge asymmetry between the particles emitted in the half-spaces below and above the reaction place, we need to know the net charge fluctuations on the freeze-out surface.
To be precise, what we need is the charge asymmetry of the particle distribution functions on the freeze-out surface. We can then use the Cooper-Frye formula to calculate the charge asymmetry of the emitted particles.
If we denote the freeze-out surface by $\Sigma_f$, the rate of emitted particles with momentum $p$  is given by 
\be
p^0\frac{dN_i}{d^3p} = \int_{\Sigma_f} d\sigma_\mu\, p^\mu f_i({\bf x},{\bf p}) \theta(\sigma_\mu p^\mu) .
\ee
The total number of particles emitted is
\be
N = \int_{\Sigma_f} d\sigma_\mu\, n^\mu({\bf x})
\ee
where 
\be 
n^\mu(x) = \int d^3p\, \frac{p^\mu}{p^0} f_i({\bf x},{\bf p}) \theta(\sigma_\mu p^\mu)
\ee
is the outward directed particle current on the freeze-out surface. The electric current and emission of net charge is obtained by multiplying each species with its electric charge $e_i$. The observable is the difference between the total charge emitted into the half-spaces above and below the collision plane. In the limit of a completely transparent medium, this observable is given by
\ba
\Delta Q &=& \int d^3p \sum_i e_i \frac{dN}{dp^3}\, {\rm sgn}(p_z)
\\
&=& \int d^3p \int_{\Sigma_f} \frac{d\sigma_\mu p^\mu}{E} \sum_i e_i f_i({\bf x},{\bf p})\,  {\rm sgn}(p_z) \theta(\sigma_\mu p^\mu) .
\nn
\ea
This quantity is related to the local charge density on the freeze-out hypersurface $\Sigma_f$, which can be expressed as
\be 
\rho(x)|_{\Sigma_f} = \int d^3p \sum_i e_i f_i({\bf x},{\bf p})|_{\Sigma_f} .
\ee

\subsubsection{Isochronous freeze-out with flow}

To obtain a first result and perform a specific calculation, we now specialize to the case of an isochronous freeze-out with transverse flow ${\bf v}({\bf x})$. This means that the entire matter volume freezes at a time $\tau=\tau_f$, and we ignore any opacity for the produced matter. Thus, fluid cells in the upper hemisphere, which have a flow vector pointing in the upward direction, will be converted to an isotropic thermal distribution of particles in the rest frame of the fluid cell. As there is no opacity, many of these 
particles will move in the negative direction perpendicular to the reaction plane. As a result, a large part of the effect of charge separation will be washed out. Obviously, this represents an underestimate of the true magnitude of the effect. 

For isochronous freeze-out we have $p^\mu d\sigma_\mu = E d^3x$ and $\theta(\sigma_\mu p^\mu) = \theta(E) = 1$. We thus obtain on the freeze-out hypersurface:
\be
\Delta Q = \int d^3x \int d^3p \sum_i e_i f_i({\bf x},{\bf p}; \tau_f) {\rm sgn}(p_z) .
\ee
In order to relate $\rho$ and $\Delta Q$, we assume that the particle distributions at freeze-out are given by thermal distributions with flow and a local chemical potential of  electric charge, $\mu_Q({\bf x})$. For simplicity, we further assume that the thermal distributions are well approximated by Boltzmann distributions, and the baryon chemical potential vanishes:
\be
f_i({\bf x},{\bf p};\tau_f) = \exp \left[ - u_\mu({\bf x}) p^\mu/T_f + e_i \mu_Q({\bf x})/T_f \right] ,
\ee
where $T_f$ is the freeze-out temperature. Since $\mu_Q({\bf x})$ is assumed to be small, we can expand to first order in this quantity. We also find that in the absence of a baryon chemical potential, particle and antiparticle distributions are equal, and thus the terms independent of $\mu_Q$ sum to zero in both, $\rho$ and $\Delta Q$. The expressions linear in $\mu_Q$ are:
\be
\rho({\bf x},\tau_f) = \mu_Q({\bf x}) \int d^3p \sum_i e_i^2 f_i^{(0)}({\bf x},{\bf p}; \tau_f) ,
\ee
and
\be
\Delta Q = \int d^3x\, \mu_Q({\bf x}) \int d^3p \sum_i e_i^2 f_i^{(0)}({\bf x},{\bf p}; \tau_f) {\rm sgn}(p_z) ,
\ee
where the superscript ``(0)'' indicates that we have set $\mu_Q=0$. We can now eliminate $\mu_Q({\bf x})$ and obtain an expression of $\Delta Q$ in terms of $\rho$:
\be
\label{DeltaQ1}
\Delta Q = \int d^3x\, \rho({\bf x},\tau_f)\, \langle {\rm sgn}(p_z) \rangle ,
\ee
where
\be
\langle {\rm sgn}(p_z) \rangle = \frac{\int d^3p \sum_i e_i^2 f_i^{(0)}({\bf x},{\bf p}; \tau_f) {\rm sgn}(p_z)}
  {\int d^3p \sum_i e_i^2 f_i^{(0)}({\bf x},{\bf p}; \tau_f)}  .
\ee
We note that the expression on the right-hand-side depends on position ${\bf x}$ only through the flow velocity ${\bf v}({\bf x})$. We also note that $\rho$ is the divergence of a vector, as given by eq.~(\ref{rho}) below. Thus, the spatial integral over $\rho$ vanishes, and $\Delta Q=0$ in the absence of transverse flow, as remarked earlier.

There are two ways of proceeding from here. The first approach assumes that the flow velocity is not large and proceeds by expanding the momentum distribution $f_i^{(0)}$ up to linear order in ${\bf v}$:
\be
f_i^{(0)}({\bf x},{\bf p}; \tau_f) \approx \exp(-E_{\bf p}/T_f) (1- {\bf v}({\bf x})\cdot{\bf p}/T_f) .
\ee
Here we neglected all terms of higher order in ${\bf v}$. Since the integrand in the numerator of the square bracket in (\ref{DeltaQ1}) contains the factor ${\rm sgn}(p_z)$, only the term proportional to $v_z$ contributes. On the other hand, the term ${\bf v}\cdot{\bf p}$ does not contribute to the integral in the denominator. We thus obtain
\begin{widetext}
\be
\Delta Q \approx \int d^3x\, \rho({\bf x},\tau_f)\, \left[ \frac{\int d^3p \sum_i e_i^2 \exp(-E_{\bf p}/T_f) 
  v_z |p_z|/T_f}{\int d^3p \sum_i e_i^2 \exp(-E_{\bf p}/T_f)} \right]
= \int d^3x\, \rho({\bf x},\tau_f)\, v_z({\bf x})\, 
\frac{\sum_i e_i^2 (6T_f^2-m_i^2)}{2 \sum_i e_i^2 m_i^2 K_2(m_i/T_f)} .
\ee
Assuming that the sum over particle species is dominated by pions, and using $m_\pi/T_f \approx 1$, we find
\be
\label{DeltaQ2}
\Delta Q \approx \frac{3}{2}  \int d^3x\, \rho({\bf x},\tau_f)\, v_z({\bf x}) .
\ee
The integral does not vanish, because both $\rho({\bf x})$ and $v_z({\bf x})$ are antisymmetric with respect to the collision plane.

In the second approach, which has the advantage that we do not need to make an assumption about the magnitude of the flow, we start by using (\ref{rho}) to replace $\rho$ with the anomalous current ${\bf j}_{\rm an}$ in (\ref{DeltaQ1}). Since both, the numerator and denominator of the term in square brackets involve an integral over all momenta, we can make a Lorentz boost into the local matter rest frame at each location. This replaces the particle distributions with those in the rest frame, $f_i({\bf x},{\bf p};\tau_f) \to \exp(-E_{\bf p}/T_f)$, the momentum integrals acquire a Lorentz factor $\gamma({\bf x})$ with $\gamma=(1-{\bf v}^2)^{-1/2}$, and the argument of the sign function transforms to $p'_z({\bf x}) = p_z+(\gamma-1)v_z\hat{\bf v}\cdot{\bf p} + \gamma v_z E_{\bf p}$, where $p'_z$ denotes the $z$-component of the momentum in the laboratory frame and $p_z$ is the $z$-component of the momentum in the comoving frame. We thus obtain the following expression for $\Delta Q$:
\begin{eqnarray}
\Delta Q &=& - \int_0^{\tau_f} dt \int d^3x\, \nabla\cdot{\bf j}_{\rm an}({\bf x},t)\, \left[
  \frac{\int d^3p \sum_i e_i^2 \exp(-E_{\bf p}/T_f)\, {\rm sgn}[p'_z({\bf x})]}
  {\int d^3p \sum_i e_i^2 \exp(-E_{\bf p}/T_f)} \right] .  
\end{eqnarray}
The ${\bf x}$ dependence of the term in brackets is contained solely in the argument of the sign function. We now integrate by parts and obtain
\be
\label{DeltaQ3}
\Delta Q = 2 \int_0^{\tau_f} dt \int d^3x\, {\bf j}_{\rm an}\cdot\nabla p'_z({\bf x})\, \left[
  \frac{\int d^3p \sum_i e_i^2 \exp(-E_{\bf p}/T_f)\, \delta[p'_z({\bf x})]}
  {\int d^3p \sum_i e_i^2 \exp(-E_{\bf p}/T_f)} \right] .
\ee
\end{widetext}
The integral in the numerator is restricted to those particles which are at rest in the direction perpendicular to the collision plane. The result (\ref{DeltaQ3}) has a simple interpretation: It counts the number of charged particles that are caused by the anomalous current to change direction from downward motion with respect to the collision plane to upward motion. The factor 2 accounts for the fact that any change in direction causes a gain in the number of upward moving charges and a loss in the downward moving charges. It is also evident from (\ref{DeltaQ3}) that the gradient of the flow velocity selects the component of the anomalous current that is parallel to it.  Thus, if there is an anomalous current in any direction other than perpendicular to the reaction plane it will  contribute to not only the charge separation perpendicular to the reaction plane but also that parallel to the reaction plane. 

In the following, we will use the form (\ref{DeltaQ2}). The fluctuation of the up-down charge asymmetry is then simply given by
\ba
\label{DeltaQQ}
\left\langle (\Delta Q)^2 \right\rangle & \approx & \frac{9}{4} \int d^3x \int d^3x' \, \langle \rho({\bf x},\tau_f) \rho({\bf x}',\tau_f)\rangle 
\nn \\
&& \qquad\qquad \times v_z({\bf x},\tau_f) v_z({\bf x}',\tau_f) .
\ea
We now must relate the correlator of the separated charge density $\rho$ to the correlator of the anomalous electric current.

As discussed above, we can neglect collective transverse flow during the phase of the collision when most of the charge separation with respect to the collision plane occurs. The induced charge density $\rho$ then satisfies the continuity equation $\dot\rho + \nabla\cdot{\bf j}_{\rm an} = 0$. We can thus calculate the charge density by integrating the continuity equation:
\be
\label{rho}
\rho({\bf x},\tau) = -\int_0^\tau dt\, \nabla\cdot{\bf j}_{\rm an}({\bf x},t) .
\ee
The charge fluctuation strength (\ref{DeltaQQ}) can thus be expressed as:
\ba
\label{DeltaQQ1}
\left\langle (\Delta Q)^2 \right\rangle & \approx & \frac{9}{4}  \int_0^{\tau_f} dt\, dt' \int d^3x \int d^3x' 
\\ && \hspace{-1.6cm}
\times \nabla v_z({\bf x},\tau_f) \cdot \langle  {\bf j}_{\rm an}({\bf x},t) {\bf j}_{\rm an}({\bf x}',t')\rangle 
\cdot  \nabla' v_z({\bf x'},\tau_f) .
\nn
\ea
We now introduce the local integrated current fluctuation strength tensor $C^{ik}$:
\be
\label{C-j}
C^{ik}({\bf x},\tau_f) = \int_0^{\tau_f} dt \, dt'  \int d^3x' \, 
\langle j^i_{\rm an}({\bf x},t) j^k_{\rm an}({\bf x}',t')\rangle .
\ee
This allows us to express the fluctuations of the charge asymmetry as
\be
\label{DeltaQ-j}
\left\langle (\Delta Q)^2 \right\rangle \approx \frac{9}{4} \int d^3x\, C^{ik}({\bf x},\tau_f)\, 
\nabla_i v_z({\bf x},\tau_f) \nabla_k v_z({\bf x},\tau_f) .
\ee

The continuity equation used to obtain the relation (\ref{rho}) does not include the effects of advection and diffusion. The actual current is composed of the anomaly induced current part and the advective part;
\be 
{\bf j}={\bf j}_{\rm an} + \rho{\bf v} .
\ee 
Charge conservation implies
\be 
\frac{\partial\rho}{\partial t} + {\bf v}\cdot{\bf \nabla}\rho = -{\bf \nabla}\cdot{\bf j}_{\rm an}.
\ee 
The effect of diffusion on the charge distribution can be included by adding the diffusion term (see e.g. \cite{Shuryak:2000pd}) in the matter frame:
\be 
-D_{\rm ch}\nabla^2 \rho .
\ee 
In any other frame, in which the matter is moving with four-velocity $u^\mu$, the diffusion term has the form
\be 
D_{\rm ch} (g^{\mu\nu}-u^\mu u^\nu)\partial_\mu \partial_\nu .
\ee 
These equations lay the ground for a future comprehensive and quantitative study of the formation of reaction plane charge asymmetry fluctuations by the action of the magnetic field on the strongly interacting matter. 

\subsubsection{Geometric approximation}

The derivation of the fluctuations of charge asymmetry leading up to Eq.~\eqref{DeltaQ-j} was carried out in the $D_{\rm ch} \rightarrow 0$ limit.  We will not pursue the consideration of a finite $D_{\rm ch}$ in much greater detail in this article. We will only consider the opposite limit of a very large diffusion coefficient $D_{\rm ch} \rightarrow \infty$, which renders the medium virtually opaque. In this limit, the 
computation of the produced charge asymmetry is greatly simplified. 

One may assume that any charge excess in the upper hemisphere of the fireball will contribute to the final charge asymmetry with respect to the reaction plane in momentum space, and that the same holds for any charge excess in the lower hemisphere, i.e.,  the large diffusion coefficient does not allow the charges to move back to the opposite hemisphere.  This assumption clearly overestimates the effect, because not all charges in the upper hemisphere will eventually be emitted with an upward component of the momentum.  However, it is useful to obtain an upper limit on the size of the expected effect which, together with the result of the calculation assuming an isochronous freeze-out, can help bracket the prediction of a more sophisticated calculation.

In this limit we define the charge asymmetry fluctuation geometrically as
\be
\langle (\Delta Q)^2 \rangle = - \int_{z<0} d^3x \; \int_{z>0} d^3x'\, \langle \rho({\bf x}) \rho({\bf x}') \rangle .
\ee
Using the continuity equation, as before, we can rewrite this as a double integral over the charge current through the reaction plane:
\be
\label{eq:DelQ2}
\langle (\Delta Q)^2 \rangle = \int d^4x\, d^4x'\, \delta(x^3) \delta(x'^3)\,  \langle j^3(x) j^3(x') \rangle .
\ee
Note that the minus has disappeared, because one of the currents flows upward through the reaction plane; the other one downward.  As already noted, in conjunction with the  $D_{\rm ch} \rightarrow 0$ calculation this geometric estimate allows us to bracket the uncertainty in our results arising from the flow and diffusion dynamics of the final state. Of course, this does not provide for an estimate of the inherent uncertainty in the approximations leading up to the calculation of the local anomalous current density.

\section{Estimate of the Charge Asymmetry Fluctuations}

In this section, we derive semi-quantitative estimates for the charge asymmetry generated by the chiral magnetic effect in collisions of two Au nuclei at the top RHIC energy. We first consider the contribution from gluon fusion as depicted in Fig.~\ref{fig-J-gg}, where the two gluons are part of the incoming 
nuclear gluon distributions. The second contribution corresponds to the case where the gluons are part 
of the thermal distribution in the deconfined phase. Finally, we consider contributions from the hadronic process depicted in Fig.~\ref{fig-tri-rho} in a thermal environment with a temperature $T\approx 160$ MeV. 

We begin with the calculation of the completely transparent medium. i.e. the $D_{\rm ch} \rightarrow 0$ 
limit, followed by the geometric estimate for  completely opaque ($D_{\rm ch} \rightarrow \infty$) medium for both partonic processes. We only derive the estimate for the upper bound of an opaque medium for the hadronic contribution. We finally note that one is really interested in the quantity
\be
\langle (\Delta (N_{+}-N_{-}))^2 \rangle = \frac{\langle (\Delta Q)^2 \rangle}{e^2} ,
\ee
because experimentally one measures the difference in the {\em number} of equally or oppositely charged particles emitted into the two hemispheres, not the charge difference.

\subsection{CGC Scenario}

\subsubsection{Anomalous current correlator}

We first consider the contribution depicted in Fig.~\ref{fig-ggetabj}, which describes the anomalous electric current generated by the fusion of two gluons from the colliding nuclei, in the pseudoscalar channel, due to the presence of the strong magnetic field carried by the colliding nuclei. We now estimate the integrated magnitude of this current and compare it with the overall quark pair multiplicity.

\begin{center}
\begin{figure}[!htb]
\includegraphics[width=0.95\linewidth]{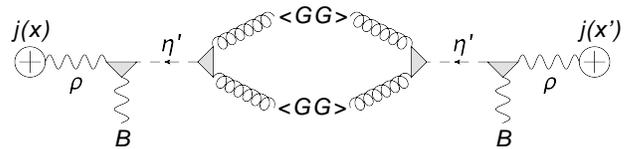}
\caption{Feynman diagram for the electromagnetic current-current correlator $\langle{\bf j}(x){\bf j}(x')\rangle$ generated by gluon fusion in the pseudoscalar channel in the presence of an external magnetic field ${\bf B}$.}
\label{fig-jj-gg}
\end{figure}
\end{center}

We start from the following expression for the current-current correlation function, shown graphically in Fig.~\ref{fig-jj-gg}:
\begin{multline}
\label{eq:jj-gg}
\langle j_z(x) j_z(x') \rangle = \left( \frac{e m_\rho^2}{g_{\rho}}\, \frac{e g_{\rho\eta'\gamma}}{m_\rho} \, 
\frac{\sqrt{3} \alpha_s \cos\theta}{2\pi f_\eta} \right)^2 
\\ 
\times 
\int d^4y d^4w d^4w' d^4y' \, \langle \rho^3(x) \rho^3(y) \rangle \, B_z(y) 
\\ 
\times
\langle \dot{\eta}'(y) \eta'(w) \rangle \, \langle ({\bf E}^a\cdot{\bf B}^a)(w) ({\bf E}^b\cdot{\bf B}^b)(w') \rangle 
\\
\times
\langle \eta'(w') \dot{\eta}'(y') \rangle \, B_z(y') \, \langle \rho^3(y') \rho^3(x') \rangle .
\end{multline}
This expression describes the process, whereby a gluon pair fuses into a virtual $\eta'$-meson, which is converted into a neutral $\rho$-meson by the intermediation of an external magnetic field. Equation (\ref{eq:jj-gg}) assumes the vector meson dominance relation (\ref{eq:VMD}) for the electromagnetic current.  

One might argue that it would be more appropriate to describe this process at the partonic level, because the final state into which the current is imbedded is a pre-equilibrium quark-gluon plasma and not a hadronic environment. However, a first estimate is more easily obtained using the effective hadronic representation of this process discussed in Sect. II.A. We expect that the magnitude of the hadronic and partonic mechanisms shown in Figs.~\ref{fig-box} and ~\ref{fig-eta-gg} is related by the parton-hadron duality property of QCD \cite{Dokshitzer:1991eq}. A confirmation of this conjecture by an explicit evaluation of the analogous partonic diagram would be desirable; we will not attempt it here. 

We now note that all elementary correlators in (\ref{eq:jj-gg}), being controlled by hadronic states with masses of the order of 1 GeV, are highly localized compared with the size of the nuclear interaction region. It thus makes sense to approximate them by space-time delta functions, in particular:
\ba
\langle \rho_z(x) \rho_z(y) \rangle & \approx & \frac{1}{m_\rho^2} \delta^{(4)}(x-y) ,
\\
\langle \dot{\eta}'(y) \eta'(w) \rangle & \approx & \frac{1}{m_{\eta'}} \delta^{(4)}(y-w) .
\ea
We also note that the magnetic field $B_z(y)$ is localized in the longitudinal direction due to Lorentz contraction of the nuclear charge distribution to a region of characteristic width $R/\gamma \approx 0.05$ fm (for a Au nucleus at top RHIC energy of 100 GeV/u). Since this is even less than the characteristic width of the gluon cloud of the color glass condensate (approximately $Q_s^{-1} \approx 0.1$ fm), we can average the magnetic field over a region of width $Q_s^{-1}$ in the longitudinal direction. Using the result derived for $B_{\rm int}$ derived in the Appendix, this yields
\be
e\bar{B}_z \approx Q_s eB_{\rm int} \approx 2 Z\alpha\, \frac{b \, \gamma}{R^3} ,
\ee
where $b$ denotes the impact parameter of the nuclear collision. We note that the average value $\bar{B}_z$ is nearly independent of the transverse coordinates within the nuclear reaction volume.

Introducing the abbreviation
\be
\label{eq:C}
C \approx \frac{g_{\rho\eta'\gamma}^2}{g_{\rho}^2}\, 
\frac{12 (Z\alpha)^2 \alpha_s^2 \cos^2\theta}{(2\pi f_\eta)^2 m_{\eta'}^2 m_\rho^2} \,
\frac{b^2  \gamma^2}{R^6} ,
\ee
we then obtain the following expression for the current-current correlator:
\be
\label{jzjz}
\langle j_z(x) j_z(x') \rangle \approx
\\
e^2 \, C \langle ({\bf E}^a\cdot{\bf B}^a)(x) ({\bf E}^b\cdot{\bf B}^b)(x') \rangle .
\ee

We now have to evaluate the correlator of the color glass condensate fields in the two colliding nuclei. Because the color fields in a fast nucleus are of the Weizs\"acker-Williams kind, i.e.\ approximately plane waves with orthogonal polarizations of the chromoelectric and chromomagnetic field strengths, the relevant contributions to the pseudoscalar invariant ${\bf E}^a\cdot{\bf B}^a$ arise when ${\bf E}^a$ and ${\bf B}^a$ originate in different nuclei:
\begin{multline}
\label{EaBaEaBa}
\left\langle [{\bf E}^a(x)\cdot{\bf B}^a(x)] [{\bf E}^b(x')\cdot{\bf B}^b(x')] \right\rangle 
\\
= \sum_{\alpha\neq\beta} \left\langle [{\bf E}^a_\alpha(x)\cdot{\bf B}^a_\beta(x)] 
  [{\bf E}^b_\alpha(x')\cdot{\bf B}^b_\beta(x')] \right\rangle 
\\
= \sum_{\alpha\neq\beta} \langle E^a_{i,\alpha}(x) E^b_{j,\alpha}(x') \rangle
 \, \langle B^a_{i,\beta}(x) B^b_{j,\beta}(x') \rangle  ,
\end{multline}
where the indices $\alpha,\beta=1,2$ count the two colliding nuclei and $i,j$ denote the spatial vector indices. 

We can relate these matrix elements to the gluon distribution function $G(\xi)$ in the colliding nuclei as follows. Color and rotational symmetry imply that for a single, fast moving nucleus the matrix element is diagonal in the indices $(a,b)$ and $(i,j)$, where $i,j$ are directions transverse to the beam, and that the chromo-electric and -magnetic correlators are equal:
\ba
\label{EaEa}
\langle E^a_i(x) E^b_j(x') \rangle &=& \langle B^a_i(x) B^b_j(x') \rangle
\nn \\
&& \hspace{-2cm} 
= \frac{\delta_{ij} \delta_{ab}}{2(N_c^2-1)}  \langle E^a_i(x) E^a_i(x') \rangle
\nn \\
&& \hspace{-2cm} 
= \frac{\delta_{ij} \delta_{ab}}{4(N_c^2-1)}  \langle [ E^a_i(x) E^a_i(x') + B^a_i(x) B^a_i(x') \rangle
\nn \\
&& \hspace{-2cm} 
= \frac{\delta_{ij} \delta_{ab}}{2(N_c^2-1)}  \langle F^{a+i}(x) F^{a+i}(x') \rangle ,
\ea
where $F^{a+i} = (E^a_i + \varepsilon_{3ij}B^a_j)/\sqrt{2}$ is the transverse chromodynamic field strength in the direction along the light cone. The gluon distribution function in a proton is defined as a light-cone Fourier transform of precisely this matrix element:
\be 
\label{eq:xG}
\xi G(\xi) = \int \frac{du^{-}}{2\pi\, 2p^+}\, e^{-\xi p^+u^{-}} \langle p | F^{a+i}(u^{-}) F^{a+i}(0) | p \rangle .
\ee
Neglecting nuclear modifications of the gluon distribution in the proton, the nuclear matrix element (\ref{EaEa}) is related to the matrix element in the proton as:
\begin{multline}
\langle F^{a+i}(x) F^{a+i}(x') \rangle_A
 \\
 = \langle p | F^{a+i}(u) F^{a+i}(0) | p \rangle \frac{\rho(\bar{\bf x}_\perp,\bar{x}^-)}{2p^+} ,
 \end{multline}
where $\bar{x}=(x+x')/2$ and $u=x-x'$, and $\rho(\bar{\bf x}_\perp,\bar{x}^-)$ is the nuclear density distribution, which is normalized to $A$. Nuclear modifications can be taken into account by using a modified gluon distribution function of the nucleon. Inverting the Fourier transform in (\ref{eq:xG}) we obtain the relation:
\begin{multline} 
\label{F+iF+i}
\langle F^{a+i}(x) F^{a+i}(x') \rangle_A
\\
= p^+ \int d\xi\, e^{-i\xi p^+ (x^--x'^-)} \, \xi G(\xi) \rho(\bar{\bf x}_\perp,\bar{x}^-) .
\end{multline}
Since we are here interested in the charge asymmetry at mid-rapidity for heavy-ion collisions in the RHIC energy domain, the relevant value of the Bjorken parameter $\xi$ of the gluon distribution function is $\xi_0 \sim 10^{-2}$ at a scale of order $Q^2 \sim m_{\eta'}^2 \approx 1 {\rm ~GeV}^2$. Approximating 
\be
\label{eq:xG0}
\xi_0 G(\xi_0)  \approx 3
\ee 
as a constant in this kinematic range, the Fourier integral evaluates to
\be
p^+ \int d\xi\, e^{-i\xi p^+ (x^--x'^-)} = 2\pi \, \delta(x^--x'^-) .
\ee
If we also approximate the Lorentz contracted nuclear density distribution as
\be
\rho(\bar{\bf x}_\perp,\bar{x}^-) \approx T_A(\bar{\bf x}_\perp)\, \delta(\bar{x}^-) ,
\ee
we obtain the desired connection between the nuclear matrix element of the gauge field strength fluctuation:
\begin{multline}
\langle F^{a+i}(x) F^{a+i}(x') \rangle_A
\\
= 2\pi\, \delta(x^-)\, \delta(x'^-) \, [\xi_0 G(\xi_0)] \, T_A(\bar{\bf x}_\perp) .
\end{multline}

Combining this result with the analogous expression for the second nucleus traveling in the negative light-cone direction, we finally obtain for the matrix element (\ref{EaBaEaBa}):
\begin{multline}
\label{EaBaEaBa-fin}
\left\langle [{\bf E}^a(x)\cdot{\bf B}^a(x)] [{\bf E}^b(x')\cdot{\bf B}^b(x')] \right\rangle 
\\
\qquad = \frac{(2\pi)^2}{N_c^2-1}\, \delta(x^-)\, \delta(x'^-)\, \delta(x^+)\, \delta(x'^+) 
\\
\times [\xi_0 G(\xi_0)]^2 \, T_{AA}(\bar{\bf x}_\perp ; {\bf b}) ,
\end{multline}
where
\be
T_{AA}({\bf x}_\perp ; {\bf b})
= T_A\left({\bf x}_\perp - \frac{\bf b}{2} \right) \, T_A\left({\bf x}_\perp + \frac{\bf b}{2} \right) .
\ee
A similar expression is obtained when one evaluates the gauge field strength correlator (\ref{EaBaEaBa}) in the color glass condensate model. Following Lappi \cite{Lappi:2006hq}, one obtains
\be
\label{xGvsCGC}
A \, [\xi_0 G(\xi_0)]  = \frac{(N_c^2-1)\, R^2 Q_s^2}{8\pi^2 \alpha_s} ,
\ee
where $Q_s(\xi_0)$ is the nuclear saturation scale and $R$ is the nuclear radius. The values (\ref{eq:xG0}) for $[\xi_0 G(\xi_0)]$, $\alpha_s = 0.3$, and $R=7$ fm for a Au nucleus correspond to a choice of the saturation scale, $Q_s^2 = 1.7 {\rm ~GeV}^2$.

\subsubsection{Isochronous freeze-out approximation}

We now make use of the isochronous freeze-out  result (\ref{DeltaQQ1}) in the $D_{\rm ch} \rightarrow 0$ limit for the final charge asymmetry fluctuation. In doing so, we apply the approximation $[{\bf B}\cdot\nabla v_z] \approx [B_z \partial_z v_z]$, which is exact for the type of transverse flow field we will be considering. The current-current correlator from Eq.~(\ref{jzjz}) can then be used to yield the charge fluctuation:
\ba
\label{eq:I-1}
\langle (\Delta Q)^2 \rangle &\approx & \frac{9}{4} e^2 C  \int d^4x\, d^4x' \, 
[\partial_z v_z(x)] [\partial'_z v_z(x')]
\nn \\
& \times & \langle ({\bf E}^a\cdot{\bf B}^a)(x) ({\bf E}^b\cdot{\bf B}^b)(x') \rangle .
\ea
We remind the reader that $y,z$ denote the transverse directions with respect to the beam in the four-vector $x^\mu=(t,x,y,z)$. Making use of the expression (\ref{EaBaEaBa-fin}) for the color field correlator, we can perform the integrals over $t,t',x,x'$ and write (\ref{eq:I-1}) as $\langle (\Delta Q)^2 \rangle = (9/4)e^2CI$ with
\ba
\label{eq:I-2}
I &=& \frac{(2\pi)^2}{N_c^2-1} \int d^2{\bf x}_\perp d^2{\bf x}_\perp' \, 
[\partial_z v_z({\bf x}_\perp)] [\partial'_z v_z({\bf x}'_\perp)] 
\nn \\
&& \times [\xi_0 G(\xi_0)]^2 \, T_{AA}(\bar{\bf x}_\perp;{\bf b}) .
\ea
The double integral can be expressed as one over $\bar{\bf x}_\perp$ and one over the difference $\Delta{\bf x}_\perp={\bf x}_\perp-{\bf x}'_\perp$. The integral over $\Delta{\bf x}_\perp$ requires some explanation. At first sight, the integrand depends on $\Delta{\bf x}_\perp$ only very weakly through the collective flow field $v_z({\bf x}_\perp)$. It turns out that this is an artifact of our use of the integrated gluon distribution function $G(\xi)$, where we expressed the nuclear matrix element (\ref{F+iF+i}) in terms of the gluon distribution. For ${\bf x}_\perp \neq {\bf x}'_\perp$ we should have used the unintegrated gluon distribution $G(\xi,{\bf k}_\perp^2)$ instead. This would have led to the replacement
\be
[\xi_0 G(\xi_0)] \longrightarrow \int \frac{d^2{\bf k}_\perp}{(2\pi)^2}\, 
[\xi G(\xi,{\bf k}_\perp^2)]\, e^{i{\bf k}_\perp\cdot\Delta{\bf x}_\perp} .
\ee
Assuming that the transverse fall-off of the nuclear unintegrated gluon distribution is controlled by the saturation scale, this substitution could be approximated by the analytic expression
\be
[\xi_0 G(\xi_0)] \longrightarrow [\xi_0 G(\xi_0)] \, e^{-Q_s|\Delta{\bf x}_\perp|} .
\ee
Neglecting the variation of the flow velocity gradient over distances of order $1/Q_s$, the integral over $\Delta{\bf x}_\perp$ can now be carried out, and we obtain:
\ba
I &=& \frac{(2\pi)^3[\xi_0 G(\xi_0)]^2}{(N_c^2-1)(2Q_s)^2} 
\nn \\
&& \times \int d^2{\bf x}_\perp \, [\partial_z v_z({\bf x}_\perp)]^2 T_{AA}(\bar{\bf x}_\perp;{\bf b}) .
\ea

To proceed further, we need to make explicit assumptions for the transverse flow profile and for the nuclear density profile. For the transverse flow velocity field at freeze-out we assume a self-similar, linear profile of the form
\be
\label{eq:vperp}
{\bf v}_\perp({\bf x}_\perp) = v_f {\bf x}_\perp/R ,
\ee
where $R$ is the nuclear radius. This choice implies a constant gradient: $\partial_z v_z = v_f/R$. Again for simplicity, we assume the nuclei to be approximately homogeneous, solid spheres with density $\rho$ and radius $R$, which gives us 
\be
T_A({\bf x}_\perp) = 2 \rho \sqrt{R^2-{\bf x}_\perp^2} ,
\ee
for the nuclear thickness function. A rough approximation for the nuclear overlap integral is found as:
\be
\int d\bar{y} \, T_{AA}(\bar{y};b) = \frac{9A^2}{8\pi^2R^2} f(b)
\ee
with
\be
f(b) \approx  1 - \frac{b^2}{R^2}\left(1-\frac{b}{4R}\right)^2  .
\ee
Putting everything together, we obtain
\be
\label{I-final}
I(b) = \frac{9\pi A^2[\xi_0 G(\xi_0)]^2v_f^2}{8(N_c^2-1)(Q_sR^2)^2}\, f(b) .
\ee
Making use of the correspondence (\ref{xGvsCGC}), the analogous expression in the color glass condensate model is
\be
\label{I-CGC}
I^{\rm (CGC)}(b) = \frac{9(N_c^2-1)v_f^2}{(8\pi)^3\alpha_s^2} \, Q_s^2 \, f(b) .
\ee

We note that (\ref{I-CGC}) contains a factor $1/\alpha_s^2$, which accounts for the nonperturbative gluon density of the color glass condensate. This factor cancels against the factor $\alpha_s^2$ in the result for the matrix element (\ref{eq:C}). The final result for the charged particle number asymmetry fluctuation, based on expression (\ref{I-final}) is thus
\ba
\langle(\Delta (N_{+}-N_{-}))^2\rangle &=& \frac{9}{4} \, C\, I(b) 
\nn \\
&&\hspace{-2cm} = \frac{243\,\pi}{8}\, \frac{g_{\rho\eta'\gamma}^2}{g_{\rho}^2}
\frac{(Z\alpha)^2 \alpha_s^2\cos^2\theta}{(2\pi f_\eta)^2 m_{\eta'}^2 m_\rho^2} 
\nn \\
&&\hspace{-1.5cm} \times \frac{v_f^2\gamma^2A^2[\xi_0 G(\xi_0)]^2}{(N_c^2-1)Q_s^2R^8}\, \frac{b^2}{R^2}\, f(b) .
\ea
Alternatively, using the color glass condensate model expression (\ref{I-CGC}), the result reads:
\ba
\label{eq:gg-final-0}
\langle(\Delta (N_{+}-N_{-}))^2\rangle &=& C\, I^{\rm (CGC)}(b) 
\nn \\
&&\hspace{-2cm} = \frac{243}{(8\pi)^3}\, \frac{g_{\rho\eta'\gamma}^2}{g_{\rho}^2}
\frac{(Z\alpha)^2 \cos^2\theta}{(2\pi f_\eta)^2 m_{\eta'}^2 m_\rho^2} 
\nn \\
&&\hspace{-1.5cm} \times (N_c^2-1)v_f^2\gamma^2 \frac{Q_s^2}{R^4}\, \frac{b^2}{R^2}\, f(b) .
\ea
Inserting the numerical values $g_{\rho\eta'\gamma}=1.31$, $g_{\rho} = 5.03$, $Q_s = 1.3$ GeV, $\gamma=100$, $Z=79$, $R=7$ fm, we obtain:
\be
\label{eq:gg-final}
\langle(\Delta (N_{+}-N_{-}))^2\rangle \approx 5 \times 10^{-5}\, v_f^2 \, \frac{b^2}{R^2}\, f(b) .
\ee

\subsubsection{Geometric approximation}

For comparison we now calculate the charge asymmetry in the geometric model in the $D_{\rm ch} \to \infty$ limit. According to our reasoning at the beginning of this section, leading to the expression (\ref{eq:DelQ2}) for the charge asymmetry fluctuation, we need to evaluate the expression 
\be
\label{eq:I-1a}
I = \int d^4x d^4x' \, \delta(z) \delta(z') \langle ({\bf E}^a\cdot{\bf B}^a)(x) ({\bf E}^b\cdot{\bf B}^b)(x') \rangle ,
\ee
where we remind the reader that $y,z$ denote the transverse directions with respect to the beam in the four-vector $x^\mu=(t,x,y,z)$. Making use of the expression (\ref{EaBaEaBa-fin}) for the color field correlator, we can write (\ref{eq:I-1a}) in the form
\ba
\label{eq:I-2a}
I = \frac{(2\pi)^2}{N_c^2-1} \int dydy' \, [\xi_0 G(\xi_0)]^2 \, T_{AA}(\bar{y};b) .
\ea
The integral over $\Delta y = y-y'$ requires some explanation. At first sight, the integrand does not depend on $\Delta y$. This is an artifact of our use of the integrated gluon distribution function $G(\xi)$ where we expressed the nuclear matrix element (\ref{F+iF+i}) in terms of the gluon distribution. For $y \neq y'$ we should have used the unintegrated gluon distribution $G(\xi,{\bf k}_\perp^2)$, which would have led to the replacement
\be
[\xi_0 G(\xi_0)] \longrightarrow \int dk_y dk_z\, [\xi G(\xi,{\bf k}_\perp^2)]\, e^{ik_y(y-y')} .
\ee
Assuming that the transverse fall-off of the nuclear unintegrated gluon distribution is controlled by the saturation scale, this could be approximated by the analytic expression
\be
[\xi_0 G(\xi_0)] \longrightarrow [\xi_0 G(\xi_0)] \, e^{-Q_s|y-y'|} .
\ee
The integral over $\Delta y$ can now be carried out, and we obtain:
\be
I = \frac{(2\pi)^2[\xi_0 G(\xi_0)]^2}{2Q_s(N_c^2-1)}  \int d\bar{y} \, T_{AA}(\bar{y};b) .
\ee
Modeling the nuclei as homogeneous, solid spheres with density $\rho$ and radius $R$, we have 
\be
T_A({\bf x}_\perp) = 2 \rho \sqrt{R^2-{\bf x}_\perp^2} ,
\ee
for the nuclear thickness function and a good approximation for the nuclear overlap integral is:
\be
\int d\bar{y} \, T_{AA}(\bar{y};b) = \frac{3A^2}{\pi^2R^3} f(b)
\ee
with
\be
f(b) \approx  1 - \frac{b^2}{R^2}\left(1-\frac{b}{4R}\right)^2  .
\ee
Putting everything together, we obtain
\be
\label{I-final-a}
I(b) = \frac{6A^2[\xi_0 G(\xi_0)]^2}{Q_sR^3(N_c^2-1)}\, f(b) .
\ee
Making use of the correspondence (\ref{xGvsCGC}), the analogous expression in the color glass condensate model is
\be
\label{I-CGC-a}
I^{\rm (CGC)}(b) = \frac{3(N_c^2-1)}{32\pi^4\alpha_s^2} \, Q_s^3R \, f(b) .
\ee

We note that (\ref{I-CGC-a}) contains a factor $1/\alpha_s^2$, which accounts for the nonperturbative gluon density of the color glass condensate. This factor cancels against the factor $\alpha_s^2$ in the result for the matrix element (\ref{eq:C}). The final result for the charge asymmetry is thus
\ba
\langle(\Delta (N_{+}-N_{-}))^2\rangle &=& C\, I(b) 
\nn \\
&=& 72\, \frac{g_{\rho\eta'\gamma}^2}{g_{\rho}^2}
\frac{(Z\alpha)^2 \alpha_s^2\cos^2\theta}{(2\pi f_\eta)^2 m_{\eta'}^2 m_\rho^2} 
\nn \\
&& \times \frac{\gamma^2A^2[\xi_0 G(\xi_0)]^2}{Q_sR^7(N_c^2-1)}\, \frac{b^2}{R^2}\, f(b) .
\ea
Alternatively, using the color glass condensate model expression (\ref{I-CGC-a}), the result is independent of the strong coupling constant:
\ba
\label{eq:gg-final-a0}
\langle(\Delta (N_{+}-N_{-}))^2\rangle &=& C\, I^{\rm (CGC)}(b) 
\nn \\
&&\hspace{-2cm} = \frac{9}{8\pi^4}\, \frac{g_{\rho\eta'\gamma}^2}{g_{\rho}^2}
\frac{(Z\alpha)^2 \cos^2\theta}{(2\pi f_\eta)^2 m_{\eta'}^2 m_\rho^2} 
\nn \\
&&\hspace{-1.5cm} \times (N_c^2-1)\gamma^2 \frac{Q_s^3}{R^3}\, \frac{b^2}{R^2}\, f(b) .
\ea
Inserting the numerical values $g_{\rho\eta'\gamma}=1.31$, $g_{\rho} = 5.03$, $Q_s = 1.3$ GeV, $Z=79$, $R=7$ fm, we obtain:
\be
\label{eq:gg-final-a}
\langle(\Delta (N_{+}-N_{-}))^2\rangle \approx 1.7 \times 10^{-3}\, \frac{b^2}{R^2}\, f(b) .
\ee

The result (\ref{eq:gg-final-a}) obtained in the geometric model is larger than the result (\ref{eq:gg-final}) found for the isochronous freeze-out by a factor $\sim 35/v_f^2$. This difference arises from an additional factor $v_f^2/RQ_s$ in the freeze-out model and constitutes a generic effect of using the flow profile function versus the geometric separation, which should occur independently of the specific mechanism underlying the anomalous current. The same factor separating the lower and upper bound of the predicted effect should thus be present, as well, for the other charge separation mechanisms considered below.

We note that the expressions (\ref{eq:gg-final-0}) and (\ref{eq:gg-final-a0}) depend sensitively on the masses of the intermediate hadronic states, $m_{\eta'}$ and $m_\rho$. If these are strongly modified by the collision environment on a time-scale of order 0.2 fm/c, which is likely the case, the magnitude of our estimate will be affected correspondingly. In order to treat such modifications realistically, a more microscopic approach will be needed.

\subsection{QGP Scenario}

We next consider the quark-gluon plasma scenario, in which fluctuations of the winding number density of the gauge field are driven by thermally assisted transitions across the so-called sphaleron barrier \cite{Klinkhamer:1984di} between vacua characterized by neighboring integer winding numbers. The thermal diffusion rate per unit volume for the Chern-Simons number in QCD, also called the ``strong sphaleron rate'', was numerically computed by Moore \cite{Moore:1997im} with the result:
\ba
\label{Gamma-sph}
\Gamma_{\rm sph} &=& \frac{1}{V\Delta t} \int d^4x\, d^4x'\,  \langle \rho_{\rm CS}(x) \rho_{\rm CS}(x') \rangle
\nn \\
&=& \frac{\alpha_s^2}{(2\pi)^2}
\int d^4x\, \langle ({\bf E}^a\cdot{\bf B}^a)(x) ({\bf E}^b\cdot{\bf B}^b)(0) \rangle
\nn \\
&\approx & 100 \, \alpha_s^5 \, T^4 ,
\ea 
with an estimated uncertainty of a factor of two. In the expression above $V$ denotes the volume and $\Delta t$ the integration time. In order to extract the winding number density correlator from this relation, we assume that the spatial and temporal correlation lengths are much shorter than the size of the volume occupied by the quark-gluon plasma and its lifetime. This assumption is well justified, as the characteristic size of winding number carrying fluctuations at high temperature is given by the magnetic length scale of order $(g^2T)^{-1}$, and the characteristic time of a saddle point transition is of order $(g^4T)^{-1}$, as discussed by Arnold {\em et al.} \cite{Arnold:1996dy}. All of these scales are of order 1 fm or less. We thus express (\ref{Gamma-sph}) in the form
\be
\label{EBEB-th}
\langle ({\bf E}^a\cdot{\bf B}^a)(x) ({\bf E}^b\cdot{\bf B}^b)(x') \rangle
\approx \delta^4(x-x') \, \frac{4\pi^2}{\alpha_s^2} \, \Gamma_{\rm sph} .
\ee

Starting from Eq.~(\ref{eq-J-gg}) we again make use of our previous result (\ref{DeltaQQ1}) for the final charge asymmetry fluctuation:
\ba
\langle (\Delta Q)^2 \rangle &\approx & \frac{9}{4} \int d^4x\, d^4x' \,  (\kappa\alpha\alpha_s)^2 
\\
& \times & [{\bf B}(x)\cdot\nabla v_z(x)] [{\bf B}(x')\cdot\nabla' v_z(x')]
\nn \\
& \times & \partial_t \partial_{t'} \langle ({\bf E}^a\cdot{\bf B}^a)(x) ({\bf E}^b\cdot{\bf B}^b)(x') \rangle .
\nn 
\ea
We now partially integrate with respect to $t$ and $t'$ and insert our explicit expression (\ref{EBEB-th}) for the pseudoscalar gluon density correlator:
\ba
\label{DQ2-th}
\langle (\Delta Q)^2 \rangle &\approx & (3\pi\alpha)^2 \int d^4x\, \kappa^2 \Gamma_{\rm sph} 
\nn \\
& \times & [\dot{\bf B}(x)\cdot\nabla v_z(x)]^2.
\ea
As we found in Section II, $\kappa \propto T^{-4}$ in the high-temperature phase of QCD, which implies that the product $(\kappa^2 \Gamma_{\rm sph})$ increases like $T^{-4}$ as the temperature falls.

We now make the assumption that a reasonable estimate of the product $[\dot{\bf B}\cdot\nabla v_z]$ is obtained by retaining only the $z$-component of the magnetic field: $[\dot{B}_z \partial_z v_z]$. The time behavior of $B_z$ at late times at the center of mass of the colliding nuclei was investigated in ref.~\cite{Kharzeev:2007jp} (see Fig.~A2). For proper times $\tau > \tau_0 = 0.5$ fm/c the results can be approximately parametrized as
\be
eB(\tau) \sim eB_0 \left(\frac{\tau_0}{\tau}\right)^{s(b)} ,
\ee
where $eB_0 \approx 300$~MeV$^2$ and the exponent $s$ depends somewhat on the impact parameter $b$, ranging from $s\approx 2$ for $b=4$~fm to $s\approx 3$ for $b=12$~fm. We further assume that the temperature changes with proper time as in the boost-invariant longitudinal expansion model:
\be
T(\tau) = T_0 \left(\frac{\tau_0}{\tau}\right)^{1/3} ,
\ee
with initial temperature $T_0 = 400$~MeV. Finally, we assume that $T_0$ and $eB_0$ are roughly constant over the transverse area; this leads clearly to an overestimate, which can easily be improved if so desired. We now insert these expressions into (\ref{DQ2-th}) and introduce coming space-time variables $(\tau,\eta,{\bf x}_\perp)$:
\ba
\frac{d\langle (\Delta Q)^2 \rangle}{d\eta} &\approx &  \frac{9}{16}\, e^2
\left(\frac{eB_0}{T_0^2}\right)^2  100\, \alpha_s^5 
\nn \\
& \times & \left(\frac{217\zeta(5)}{360 \pi^4}\right)^2 \int d^2x_\perp \, (\partial_z v_z)^2 
\nn \\
& \times & \frac{s^2}{\tau_0^2} \int_{\tau_0}^{\tau_h} \tau d\tau \left(\frac{\tau_0}{\tau}\right)^{2s+\frac{2}{3}} ,
\ea
where $\tau_h$ denotes the hadronization time of the quark-gluon plasma. We note that $v_z$, according to (\ref{DeltaQQ1}), is to be evaluated at the freeze-out time $\tau_f$. Dividing both sides by $e^2$ and inserting numbers we obtain
\ba
\frac{d\langle (\Delta (N_{+}-N_{-}))^2 \rangle}{d\eta} &\approx & 2\times 10^{-11} \, \frac{s^2}{2s-\frac{1}{3}}
\nn \\
& \times & \int d^2x_\perp \, (\partial_z v_z)^2 ,
\ea
where we have taken the upper limit of the integral to infinity.

To proceed further, we again assume that the transverse flow velocity field at freeze-out can be approximated by the linear self-similar linear profile (\ref{eq:vperp}). This choice implies $\partial_z v_z = v_f/R$ and thus
\be 
\label{dzvz2}
\int d^2x_\perp \, (\partial_z v_z)^2 = \pi\, v_f^2 .
\ee
Further neglecting the constant $(-1/3)$ in the denominator, we obtain out final result for the charged particle number asymmetry fluctuations with respect to the reaction plane in the QGP scenario:
\be
\frac{d\langle (\Delta (N_{+}-N_{-}))^2 \rangle}{d\eta} \approx  3 \times 10^{-11} \, s(b) \, v_f^2 .
\ee

We finally note that when one evaluates the CGC scenario in the geometric approximation, one obtains an ill-defined product of three delta functions in the $z$-direction: $\delta(z)\delta(z')\delta(z-z')$. This means that one cannot neglect the correlation length of the Chern-Simons number density in the $z$-direction, but must assign it a nonzero value $\lambda\sim (g^2T)^{-1}$. The delta function product then becomes $\delta(z)\delta(z')/\lambda$, and the final result acquires a factor $(R/\lambda)$ instead of the factor $v_f^2$.

\subsection{Corona Scenario}

We finally consider the corona scenario, where the anomalous current is generated by the hadronic process shown in diagram (\ref{fig-tri-rho}). Although we will, for definiteness, evaluate the current-current correlator in a thermal ensemble, we do not insist that this is an excellent approximation at the relevant early times when the magnetic field achieves is maximal strength. Nevertheless, we hope that the result obtained in this way gives an upper estimate of the magnitude of the expected effect. Denoting the location of the $\rho\pi\pi$ vertex by $y$ and that of the $\rho B\pi^0$ vertex by $z$, the current operator is given by
\ba
\label{j-pi-rho}
j^\mu(x) &=& i g_{\rho\pi\pi}  \Big( \Phi^*(x)\partial^\mu\Phi(x) - (\partial^\mu\Phi^*(x))\Phi(x) \Big)
\nn \\
& \times &
\int d^4y\,  \Big( \Phi^*(y)\partial^\nu\Phi(y) - (\partial^\nu\Phi^*(y))\Phi(y) \Big)  \rho_\nu(y) 
\nn \\
& \times &
\frac{e g_{\rho\pi\gamma}}{2 m_\rho} \varepsilon_{\sigma\rho\lambda\delta}
\int d^4z\, \rho^\lambda(z) \partial^\delta \phi(z) F^{\rho\sigma}(z) .
\ea
We recall that $\Phi,\phi$ denote the charged and neutral pion fields, respectively.

Making us of the translation invariance for the thermal current-current correlator, we can simplify the expression (\ref{eq:DelQ2}) for the charge asymmetry fluctuation to read
\be
\langle (\Delta Q)^2 \rangle = \int d^4x\, d^4x'\, \delta(x^3) \delta(x'^3)\,  \langle j^3(x) j^3(0) \rangle ,
\ee
which will serve as the starting point of our calculation. Translation invariance also makes it convenient to consider the Fourier transform $j^\mu(k)$ of the current. In the thermal ensemble, the Fourier transform of the current-current correlator is given by the ``$<$'' unordered correlator, because we are interested in the effect due to an incoming neutral pion and not in vacuum fluctuations. We define:
\be
\Pi^{33}(k) = \int d^4x\, e^{-ikx}  \langle j^3(x) j^3(0) \rangle .
\ee
We note that only the longitudinal component of the current vector, and thus only the longitudinal component of the $\rho$-meson field contribute to the matrix element, because we have used the continuity equation to relate the charge density to the divergence of the current. Furthermore, only the component $F^{12} = B_z$ of the magnetic field perpendicular to the reaction plane contributes to the correlator. For moment, we neglect the time- and space-dependence of the magnetic field ${\bf B}$. We also note that the integration over the coordinates ($x$ and $y$) spanning the reaction plane reduce the momentum range of interest to $k_x=k_y=0$ and $k^0 = E_\pi({\bf k}) \equiv \sqrt{k_z^2 + m_\pi^2}$.

After some algebra, one then obtains the following expression for $\Pi^{33}(k)$:
\ba
\label{Pi33}
\Pi^{33}(k) &=& g_{\rho\pi\pi}^2 \, \frac{e^2 g_{\rho\pi\gamma}^2}{m_\rho^2} \, B_z^2 \,
\frac{\pi n_\pi(k^0)}{E_\pi({\bf k})} 
\nn \\
&& \times
\frac{\tilde\Pi(k)^2}{\big( k^2-m_\rho^2-\Pi_{\rho}(k) \big)^2} ,
\ea
where $\Pi_{\rho}(k)$ is the $\rho$-meson self-energy and
\ba
\label{tildePi}
\tilde{\Pi}(k) &=& \int \frac{d^4p}{(2\pi)^4}\, \frac{2(2p^3-k^3)(p^0k^3-p^3k^0)}
{(p_0)^2-(E_\pi^{(\pm)}({\bf p}))^2}
\nn \\  
& \times  &
\frac{n(p^0)}{\big[ (p_0-E_\pi({\bf p}))^2-(E_\pi^{(\pm)}({\bf p}-\bf{k}))^2 \big]} .
\ea
Here $E_\pi^{(\pm)}({\bf p})$ denotes the (in-medium) on-shell energy of a charged pion, while $E_\pi({\bf k})$ denotes the on-shell energy of the neutral pion initiating the current. Note that the integral over $p^0$ in the last equation accounts for the sum over Matsubara frequencies. The integral over the four poles of the integrand in (\ref{tildePi}) yields two pairs of identical residues, leaving the expression:
\ba
\tilde{\Pi}(k) &=& 2 \int \frac{d^3p}{(2\pi)^3}\, \frac{n({\bf p})}{E({\bf p})} \,
\big[ E({\bf p}) k^3 - p^3 E({\bf k}) \big]
\nn \\ 
& \times &
\left[ \frac{2p^3 - k^3}{\big(E({\bf k}) - E({\bf p})\big)^2 - E({\bf p}-{\bf k})^2}\right.
\nn \\
& + & \left. 
 \frac{2p^3 + k^3}{\big(E({\bf k}) + E({\bf p})\big)^2 - E({\bf p}+{\bf k})^2}\right] ,
\ea
where we have dropped the sub- and superscripts on the pion on-shell energies to reduce the cluttering of the equation. Since the only non-zero component of ${\bf k}$ is $k^3$, we can go to polar coordinates in the integral over ${\bf p}$ and integrate over the azimuthal angle. This leaves a two-dimensional integral over $p=|{\bf p}|$ and the polar angle $\cos\theta$. This integral simplifies considerably if we neglect the pion mass relative to the temperature $T$. Since the main contribution to the anomalous current comes at early times, this may provide for a reasonable approximation. The final result is:
\ba
\label{tildePi2}
\tilde{\Pi}(k) &\approx & \frac{1}{2\pi^2} \int_0^\infty dp\, p^2 \int_{-1}^{1} d\cos\theta \, \frac{k_z}{p}\, n(p)
\nn \\
&=& \frac{1}{6} k_z T^2 ,
\ea
where we have written $k_z \equiv k^3$ for clarity.

We now need to substitute (\ref{tildePi2}) into Eq.~(\ref{Pi33}) and transform back to coordinate space. However, doing so, we need to take into account the time dependence of the magnetic field. The full expression to be evaluated is:
\ba
\langle (\Delta Q)^2 \rangle &=&  \frac{e^2 g_{\rho\pi\gamma}^2}{m_\rho^2}\,
\frac{g_{\rho\pi\pi}^2 \,}{(m_\rho^2-m_\pi^2)^2}\, \left(\frac{T^2}{6}\right)^2 
\nn \\
& \times & 
\int dt\, dx\, dy\, \int_{-\infty}^{\infty} dt' \int_0^\infty \frac{dk}{2\pi}\, e^{ikt'}
\nn \\
& \times & 
\frac{k}{2} n(k) B_z(x+\frac{x'}{2}) B_z(x-\frac{x'}{2}) .
\ea
We will use the following approximation for the magnetic field generated by a single nucleus (see (\ref{Bz}) in the ultrarelativistic limit $v \approx 1$):
\be
B_z^{(\pm)}(x,y,z,t) \approx \frac{Zeb\gamma}{4\pi} 
\frac{\theta(R-y)\theta(R-z)}{\big( R^2 + \gamma^2 (x\mp t)^2 \big)^{3/2}} ,
\ee
where the $-(+)$ sign is for a right-moving (left-moving) nucleus. The magnetic field of both nuclei is given by $B_z = B_z^{(+)} B_z^{(-)}$. 

We begin with the Fourier integral, which is approximately of the form
\ba
I_1(k) &\approx & 
\int_{-\infty}^{\infty} dt' \frac{e^{ikt'}}{\left( R^2 + \gamma^2 t'^2\right)^3}
\nn \\
&=& \frac{\pi\, e^{-kR/\gamma}}{16 \gamma R^5} 
\left[ 3 \left( \frac{kR}{\gamma} +1 \right) + \left(\frac{kR}{\gamma}\right)^2 \right] .
\ea
Next we perform the integration over the current momentum $k$. Noting that $\gamma/R \gg T$ for $\gamma = 100$ and $T \approx 150$ MeV, implying that $kR/\gamma \ll 1$ where the remainder of the integrand is large, the integral simplifies to:
\ba
I_2 &=& \int_0^\infty \frac{dk}{2\pi}\, \frac{k}{2} \left( e^{k/T}-1 \right)^{-1} I_1(k) 
\nn \\
&\approx & \frac{3}{64 R^5} \int_0^\infty dk\, \frac{k}{e^{k/T}-1}
= \frac{\pi^2 T^2}{128\gamma R^5} .
\ea
Next, we perform the integrals over $t,x,y$. Since the magnetic field depends on $\gamma(x\mp t)$, the nuclear volume is Lorentz contracted in at least one light-cone direction; a reasonable estimate may thus be
\be
\int dt\, dx\, dy \approx \frac{R^3}{\gamma} .
\ee
Collecting all factors, we get:
\be
\langle (\Delta Q)^2 \rangle \approx  \frac{(\pi Z\alpha g_{\rho\pi\gamma} g_{\rho\pi\pi})^2}{768}
\left(\frac{T}{m_\rho}\right)^6 \frac{b^2}{R^2} .
\ee
Inserting $Z=79$, $g_{\rho\pi\gamma} = 0.6$, $g_{\rho\pi\gamma}= 5.92$, $T \approx 150$ MeV, we finally obtain
\be
\langle(\Delta (N_{+}-N_{-}))^2\rangle = \frac{\langle (\Delta Q)^2 \rangle}{e^2} 
\approx 3 \times 10^{-5}\, \frac{b^2}{R^2} ,
\ee
which is approximately 50 times smaller than our geometric estimate (\ref{eq:gg-final-a}) for the contribution due to the color glass condensate scenario. We note once more that our result likely constitutes an overestimate of the true magnitude of the corona contribution, because a thermal pion gas will not be formed until some time (maybe 1 fm/c) after the onset of the collision, when the magnetic field strength has already subsided significantly. We did not take this delay time into consideration in our estimate.

\section{Summary}

In this work, we have explored possible sources of electric charge asymmetry fluctuations caused by the interaction of the collisional magnetic field with the highly excited QCD matter created in a relativistic heavy ion collision. We first analyzed the various microscopic mechanisms that contribute to the anomalous current and showed that the chiral magnetic effect, i.e.\ charge separation along the direction of the magnetic effect, occurs in both, the partonic and the hadronic, phases of QCD. In the partonic phase after thermalization, the anomalous current is dominantly produced by winding number carrying gauge field configurations interacting with the magnetic field via a thermal quark loop. Before thermalization, the winding number carrying gauge field configurations can interact with the magnetic field via a virtual $eta'$ (or $\eta$) meson, which electromagnetically converts into a $\rho$-meson. In the thermal hadronic phase, the anomalous current is predominantly generated by the electromagnetic $\pi-\rho$ conversion process $\pi^{0} + \gamma \to \rho^{0} \to \pi^{+} +\pi^{-}$.

It is evident from our analysis that, contrary to previous claims, reaction plane charge asymmetry fluctuations do not require ``local'' or global parity violation for their formation, as the constituent processes of these various mechanisms -- except the thermal winding number fluctuations in the quark-gluon plasma phase -- are well known from ordinary hadronic physics. We discussed five distinct scenarios for the creation of a reaction plane charge asymmetry fluctuation, viz.\ the color glass condensate (CGC) scenario, the quark-gluon plasma (QGP) scenario, the glasma scenario, the corona scenario, and the hadron gas (HG) scenario. We analyzed two of these, the CGC and the corona scenario, in some detail and found that the estimated magnitude of the expected effect in both scenarios is much smaller than the effect observed by the STAR experiment. This suggests that the observations may be due to some other process.

Rigorous predictions of the reaction plane charge asymmetry fluctuations will need to include in-medium effects on the interactions involved in the various mechanisms, such as modifications of the meson masses and effective couplings. They will also need to track the charge density transport from the creation by the anomalous current through the expanding matter up to the freeze-out hypersurface. We showed how the final charge asymmetry fluctuation can be calculated from the fluctuation of the charge density at freeze-out.  

While it is difficult to quantitatively ascertain the relative magnitude of the influence of these effects, it is reasonable to expect that in-medium modifications of meson propagators may somewhat enhance the charge asymmetry, whereas dissipative transport mechanisms active during the expansion phase will likely suppress the asymmetry. It is instructive to attempt a very rough estimate of upper bound by which in-medium modifications of the masses of the intermediate states involved in the charge asymmetry creation could enhance the expected effect. As an example, we consider the CGC scenario. Here the magnitude of the charge asymmetry fluctuation is controlled by the coefficient $C$, eq.~(\ref{eq:C}), which involves the factor $(2\pi f_\eta\, m_{\eta'}\, m_\rho)^2$ in the denominator. These factors represent scales relating to the properties of intermediate quark-anti-quark states. In the normal QCD vacuum, these will be limited below by the QCD confinements scale $\Lambda_{\rm QCD} \approx 200$ MeV; in a thermal medium, they will be limited below by the thermal mass scale of order $gT$, which is also of the order of 200 MeV or more under conditions reached at RHIC. An optimistic upper limit to the possible enhancement due to in-medium modification of the effective mass scale of quark-antiquark excitations would thus be $(2\pi f_\eta\, m_{\eta'}\, m_\rho)^2/(200~{\rm MeV})^6 \approx 10^4$.

It is thus not inconceivable that QCD processes occurring in a high energy density environment could enhance the charge asymmetry fluctuations by several orders of magnitude above our estimates. In order to result in such an extreme enhancement of the charge asymmetry fluctuations, however, the processes responsible for it must take hold on a time scale less than 1 fm/c so that they can act during the period of peak or near-peak magnetic fields. Furthermore, if the chiral magnetic effect is experimentally confirmed as the source of the observed event-by-event fluctuations, it thus would imply the experimental observation of a form of  highly excited QCD matter in which the creation of local winding number fluctuations is strongly enhanced compared to the normal QCD vacuum. 

\section*{Acknowledgments}

This work was supported in part by grants from the U.~S.~Department of Energy (DE-FG02-05ER41367 and DE-FG02-01ER41190) and from the Japanese Ministry of Education. We thank K~Fukushima and S.~Voloshin for helpful comments and H.~Warringa for pointing out a typographical error in eq.~(\ref{eq-QED-an}).

\appendix

\section{Magnetic field estimate}

The magnetic field of the two colliding nuclei is calculated from the retarded vector potential
\ba
{\bf A}(x) &=& \int D_{\rm ret}(x-x') {\bf j}_N(x') 
\nn \\
&=& \int D_{\rm ret}(x-x') \rho_N(x'-vt') \gamma{\bf v} ,
\ea
where ${\bf v}$ is the velocity of the colliding nucleus, $\gamma$ the Lorentz factor, and $\rho_N$ the static nuclear charge density. Here we have neglected any ``slow-down'' of the valence quarks as the two nuclei collide. In a more detailed calculation one would want to use a more detailed treatment of the dynamics of the matter in the fragmentation regions of the colliding nuclei.

Suppose a point charge $e$ is moving along the $x$-axis and its $x$-coordinate is given by $x=vt$. See Fig.~\ref{react-plane} for a schematic view of the kinematics and the coordinate system. The magnetic field
${\bf B}$ at ${\bf r}=(x,~y,~z)$ is given by
\be
{\bf B}=\frac{\gamma e}{r'^3}{\bf v}\times{\bf r}',
\ee
where
\be
{\bf r}'= (\gamma(x-vt),~y,~z).
\ee
When a point charge $e$ is moving along $y=\pm\frac{b}{2}$, the magnetic field created by the charge is given by
\be
{\bf B}^{(\pm)}=\frac{e\gamma}{4\pi {r'_\pm}^3}\vec{v}\times{\bf r}'_\pm,
\ee
where
\be
{\bf r}'_\pm= (\gamma(x-vt),~y\mp\frac{b}{2},~z).
\ee
In the following, we set $x=0$ and consider the magnetic field as a function of $y$ and $z$. The $z$-component of the magnetic field is the superposition of the $z$-components of the magnetic fields created by the two point charges:
\be
B_z = B^{(+)}_{z} + B^{(-)}_{z} ,
\ee
where
\be
B^{(\pm)}_{z} =  \frac{ev\gamma}{4\pi} \frac{ \frac{b}{2} \mp y}
  {\left[ \left(y\mp\frac{b}{2} \right)^2 + z^2 + v^2 \gamma^2 t^2 \right]^{3/2}} .
\label{point}
\ee

For nuclei, we adopt the approximation that the charge $Ze$ is distributed uniformly with a radius $R$.
Then the above formula (\ref{point}) is replaced by
\be
\label{Bz}
eB_z^{(\pm)}  = \left\{
\begin{array}{lr}
Z\alpha v\gamma \left(\frac{b}{2}\mp y\right) \frac{\tilde{r}_\pm}{R^4} &
\quad (\tilde{r}_\pm \le R), \cr 
& \cr
Z\alpha v\gamma \left(\frac{b}{2}\mp y\right) \frac{1}{\tilde{r}_\pm^3} &
\quad (\tilde{r}_\pm > R),
\end{array}
\right.
\ee
where
\be
\tilde{r}_\pm = \left [ \left ( y\mp\frac{b}{2} \right )^2 + z^2 + v^2 \gamma^2 t^2 \right ]^{1/2} .
\ee

\begin{center}
\begin{figure}[!htb]
\includegraphics[width=0.9\linewidth]{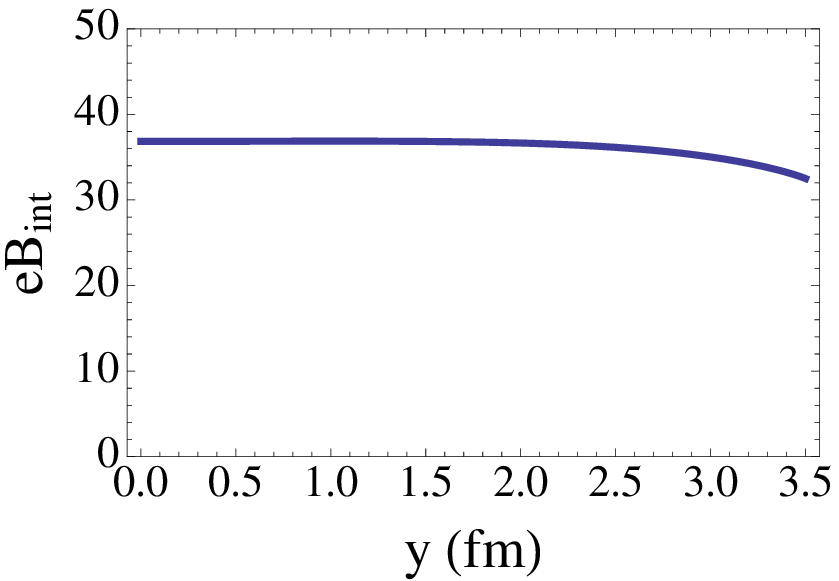}
\includegraphics[width=0.9\linewidth]{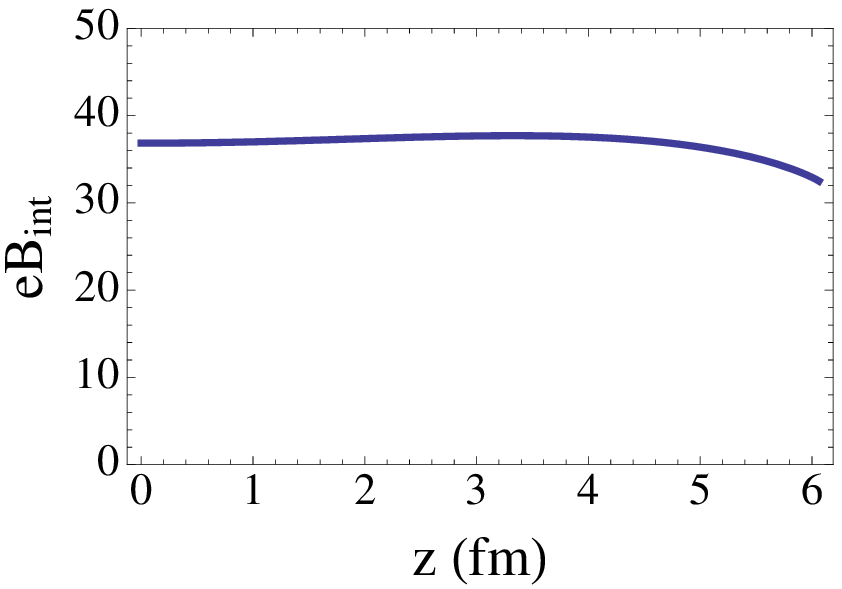}
\caption{Time-integrated magnetic field strength $eB_{\rm int}(y,z=0)$ (top panel) and $eB_{\rm int}(y=0,z)$ (bottom panel) in units of MeV for the parameters $b=7$ fm, $R = 7$ fm, and $Z=79$.}
\label{fig-Bint}
\end{figure}
\end{center}

\begin{center}
\begin{figure}[!htb]
\includegraphics[width=0.9\linewidth]{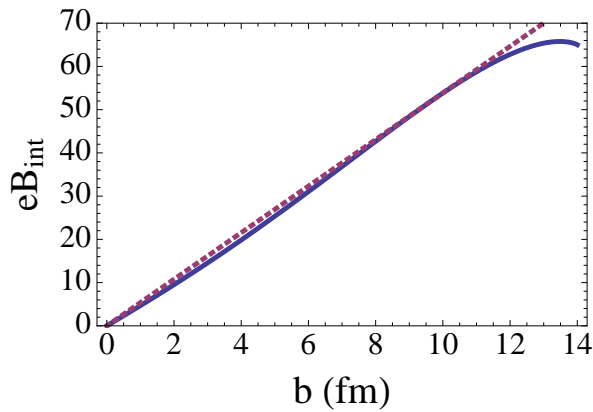}
\caption{Time-integrated magnetic field strength $eB_{\rm int}$ in units of MeV at $y=z=0$ as function of the impact parameter $b$ for the parameters $R = 7$ fm and $Z=79$. The dotted line shows the approximation (\ref{Bint-app}) for comparison.}
\label{fig-Bint-b}
\end{figure}
\end{center}

Next we calculate the time integral of $B_z$:
\be
\label{Bint}
eB_{\rm int}(y,z) = \int_{-\infty}^\infty dt\, eB_z(x=0,y,z,t) 
\ee
and note that the result is independent of the product $\gamma v$. As typical values applying to a mid-central Au+Au collision at RHIC, we take $b=7$ fm, $R = 7$ fm, and $Z=79$. The time-integrated field $eB_{\rm int}$ is shown in Fig.~\ref{fig-Bint} as function of the position $z$ perpendicular to the reaction plane for $y=0$ (top panel), and as function of the position $y$ within the reaction plane for $z=0$. As the figure shows, the time-integrated magnetic field is nearly independent of the location within the overlap region of the two nuclei. The impact parameter dependence of the time-integrated magnetic field at the center ($y=z=0$) of the transverse plane is shown in Fig.~\ref{fig-Bint-b}, in comparison with the approximate formula
\be
\label{Bint-app}
eB_{\rm int} \approx 2.32\, Z\alpha \, \frac{b}{R^2} .
\ee

\end{document}